\documentclass[openacc]{rsproca_new}%%%%where rsproca is the template name

\usepackage{graphicx}
\usepackage{epstopdf}
\usepackage{epsfig}
\usepackage{amssymb,amsmath,stmaryrd,tabularx}
\usepackage[utf8]{inputenc}
\usepackage[center]{subfigure}
\usepackage{amssymb}
\usepackage{cleveref}
\usepackage[dvipsnames]{xcolor}
\usepackage[normalem]{ulem}
\usepackage{bm}
\usepackage{appendix}
\usepackage{amsmath}
\usepackage{empheq}
\usepackage{url}
\usepackage{amsmath}
\usepackage{empheq}
\usepackage[theorems,skins]{tcolorbox}
\newtcolorbox{mymathbox}[1][]{colback=white, sharp corners, #1}
\newtcbox{\othermathbox}[1][]{nobeforeafter, math upper, tcbox raise base, enhanced, sharp corners, colback=black!10, colframe=red!30!black, drop fuzzy shadow, left=1em, top=0.5em, right=2em, bottom=0.5em}

\newcommand{\bea}{\begin{eqnarray}}  
\newcommand{\eea}{\end{eqnarray}}

\newcommand{\fet}[1]{\mathbf{#1}}

\begin{document}

%%%% Article title to be placed here
\title{Flow around topological defects in active nematic films}

\author{Jonas R\o nning$^1$, M.~Cristina Marchetti $^2$, Mark J. Bowick$^3$ and  Luiza Angheluta$^1$}

\address{$^1$Njord Centre, Department of Physics, University of Oslo, P. O. Box 1048, 0316 Oslo, Norway\\
$^2$Department of Physics, University of California Santa Barbara, Santa Barbara, CA 93106, USA\\
$^3$Kavli Institute for Theoretical Physics, University of California Santa Barbara, Santa Barbara, CA 93106, USA}

%%%% Subject entries to be placed here %%%%
\subject{soft matter, biophysics, fluid mechanics}

%%%% Keyword entries to be placed here %%%%
\keywords{active nematics, topological defects, nematic liquid crystals, hydrodynamics}

%%%% Insert corresponding author and its email address}
\corres{Jonas Rønning\\
\email{jonasron@uio.no}}

%%%% Abstract text to be placed here %%%%%%%%%%%%
\begin{abstract}
We study the active flow around isolated defects and the self-propulsion velocity of $+1/2$ defects in an active nematic film with both viscous dissipation (with viscosity $\eta$) and frictional damping $\Gamma$ with a substrate. The interplay between these two dissipation mechanisms is controlled by the hydrodynamic dissipation length $\ell_d=\sqrt{\eta/\Gamma}$ that screens the flows. For an isolated defect, in the absence of  screening  from other defects, the size of the shear vorticity around the defect is controlled by the system size $R$. In the presence of friction that leads to a finite value of $\ell_d$, the vorticity field decays to zero on the lengthscales larger than $\ell_d$. We show that the self-propulsion velocity of $+1/2$ defects grows with $R$ in small systems where $R<\ell_d$, while in the infinite system limit or when $R\gg \ell_d$, it approaches a constant value determined by $\ell_d$.    
\end{abstract}
%%%%%%%%%%%%%%%%%%%%%%%%%%%

\begin{fmtext}
\section{Introduction} \label{sec:intro}

Active matter consists of collections of individuals that dissipate energy taken from the environment to generate  motion and forces and self-organize into a rich variety of  ordered phases. Many active systems exhibit nematic order interrupted by orientational defects and advected by spontaneous flows driven by intrinsic activity of the self-propelled individuals. This behavior is found in reconstituted systems, such as mixtures of cytoskeletal filaments and motor proteins ~\cite{sanchez2012spontaneous,guillamat2017taming,needleman2017active,kumar2018tunable}, bacterial suspensions~\cite{Doostmohammadi2018, marchetti2013hydrodynamics} and cell sheets~\cite{saw2018biological,mueller2019emergence}, as well as synthetic systems, like vertically vibrated layers of granular rods \cite{kudrolli2008swarming,marchetti2013hydrodynamics}. 
\end{fmtext}
\maketitle
%%%%%%% end of first page %%%%%%%%%%%%%%
A central feature of active nematics is the feedback between active stresses which distort orientational order and the spontaneous flow generated by such distortions. In hydrodynamic descriptions~\cite{marchetti2013hydrodynamics}, the active stress $\sigma_{ij}^{a}$ exerted by elongated active entities on the surrounding fluid  is proportional to the nematic order parameter tensor $Q_{ij}$, namely $\sigma^{a}_{ij} = \alpha_0 Q_{ij}$~\cite{simha2002hydrodynamic, juelicher2007active}. The activity coefficient $\alpha_0$ embodies the microscale  biomolecular processes that convert chemical energy into mechanical forces, and depends on the concentration of active entities, which in general may vary in space and time~\cite{Giomi2013,lemma2019statistical}. The sign of $\alpha_0$ distinguishes  between contractile ($\alpha_0>0$) stress generated by ``puller" swimmers, such as the algae Chlamydomonas, versus extensile ($\alpha_0<0$) stress generated by ``pusher" swimmers, e.g., most flagellated bacteria. Its magnitude controls the strength of the active flow. Fluctuations in orientational order yield active stresses and associated flows, which can in turn enhance the orientational distortions. The resulting  feedback loop  destabilizes the nematic order, driving the system to a state of self-sustained spatio-temporally chaotic flow, with proliferation of  topological defects, and termed active turbulence~\cite{thampi2014vorticity,doostmohammadi2017onset}.

The lowest-energy orientational defects in nematic films have half-integer topological charge and opposite sign. The $+1/2$ defects have comet-like shape, whilst the $-1/2$ defects have a tri-fold symmetry (see Figs.~\ref{fig:Positive_velocity_low_damping} and \ref{fig:neg-high}). Defects strongly disrupt orientational order and induce long-range nematic distortions. In active systems, such distortions generate flows with symmetry and profiles controlled by the defect geometry. The nematic distortion created by a $+1/2$ defect yields an active flow that is finite at the defect core. A $+1/2$ defect then rides along with the flow it itself generates, behaving like  a motile particle 
with a non-vanishing self-propulsion velocity $\fet v^{a}_+$, even in the absence of external drive~\cite{Giomi2013,Pismen2013}. On the other hand, the active backflow generated by a $-1/2$ defect  vanishes at the core due to the defect's threefold symmetry (see Fig.~\ref{fig:neg-high}). Thus $-1/2$ defects behave like passive particles and have no spontaneous motility in the absence of external driving. A simple estimate demonstrates that $\fet v^{a}_+$ is directed along the polar axis of the $+1/2$ defect and is proportional to the activity $\alpha_0$. In an extensile medium $+1/2$ defects  self-propel in the direction of the head of the comet, while in a contractile system they move towards the comet's tail~\cite{Giomi2013,Pismen2013,giomi2014defect}. The direction of motion of $+1/2$ defects can then be used as a metric for determining the nature of active stress in the system. Such measurements have for instance revealed the surprising dominance of extensile stresses in confluent tissue composed of tightly bound contractile individual cells~\cite{Kawaguchi2017,Saw2017,balasubramaniam2020nature,vafa2021fluctuations,killeen2021polar}. 

The flow generated by defects and the resulting propulsive speed of the $+1/2$ also vary depending on the dissipative processes at play in the system and the role of fluid incompressibility. Specifically, important differences exist between ``dry'' systems, where dissipation is dominated by friction $\Gamma$ with a substrate or an external medium~\cite{shankar2018defect, angheluta2021role} and ``wet'' systems where dissipation is mainly controlled by viscosity $\eta$, resulting in long-range hydrodynamic effects~\cite{giomi2014defect,pismen2017viscous,thampi2014vorticity, angheluta2021role}. In incompressible wet systems, activity is also a source of pressure gradients, which in turn contribute alongside with the nematic distortion to the self-motility of positive defects. 
In the limit of viscous dominated flows with no friction with the substrate, the self-propulsion speed scales  as $|\fet v^{a}_+|\sim\frac{|\alpha_0|}{\eta} \ell$,  where $\ell$ is a length scale given by the system size for an isolated defect ~\cite{giomi2014defect} or by the mean separation between defects, which is, in turn,  controlled by the active length scale $\ell_a=\sqrt{K/|\alpha_0|}$, with $K$ the nematic stiffness~\cite{giomi2014defect}.
In overdamped (dry) systems, where viscosity is negligible compared to frictional damping with the substrate, $|\fet v^{a}_+|\sim|\alpha_0|/(\xi\Gamma)$, where $\xi$ is the nematic coherence length \cite{shankar2018defect,shankar2019hydrodynamics, angheluta2021role}. A complete calculation of the active flows associated with defect configurations and of the propulsive speed of the $+1/2$ defect that bridges between the two limits is, however, not available. The need for such a calculation is further motivated by recent work that has shown that  tuning frictional damping relative to viscous dissipation leads to different dynamical regimes and ordering behavior of interacting defects \cite{thijssen2020role}. 
 
\begin{figure}[t]
    \centering
    \includegraphics[width = 0.45\textwidth]{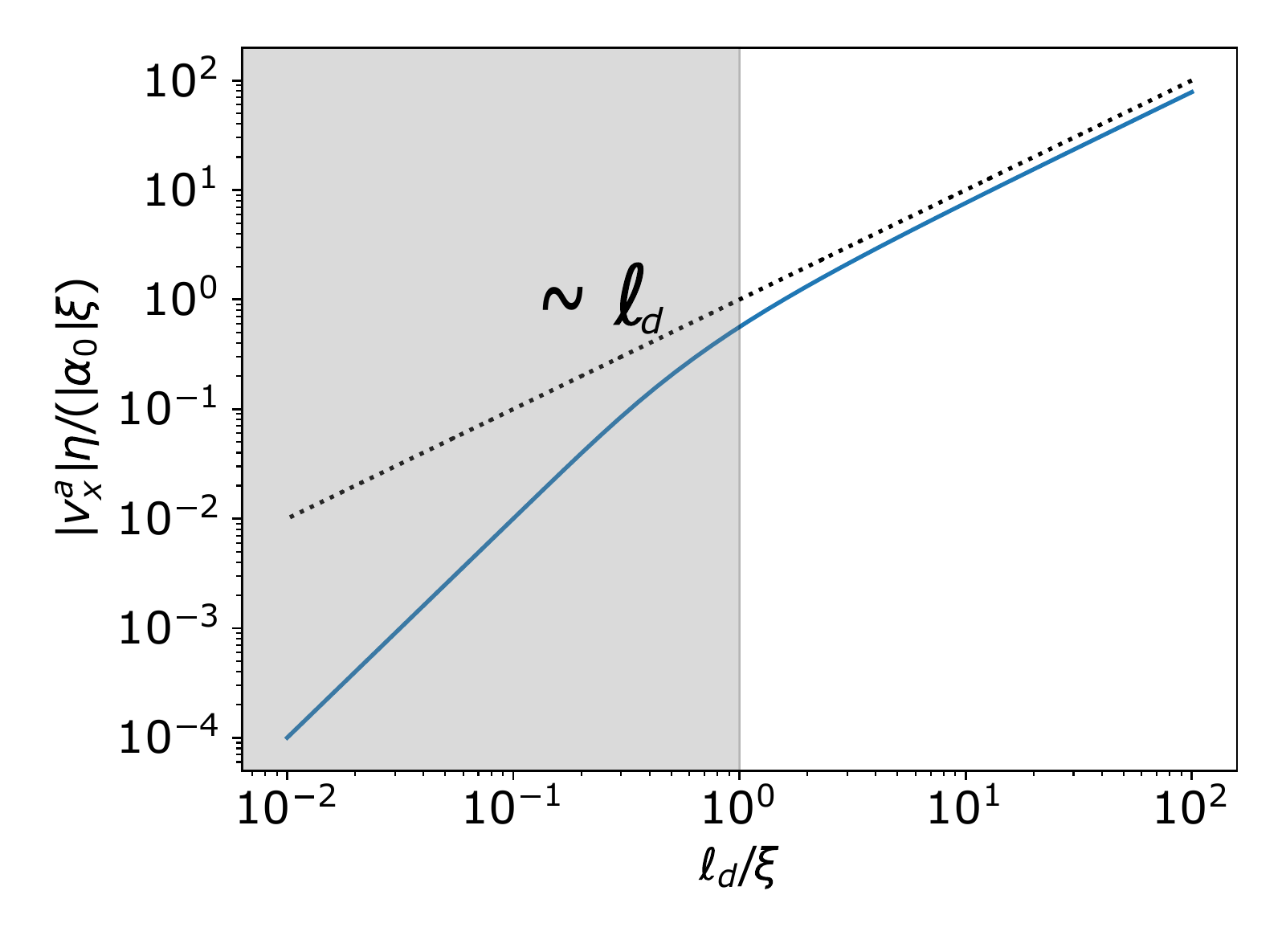}
    \caption{The self-propulsion speed of a $+1/2$ defect as function of $\ell_d/\xi$ in an unbounded system. The line is the exact analytical solution, while the dotted line shows the asymptotic scaling with $\ell_d$, i.e. $v_x^a \eta \sim \ell_d$ for $\ell_d >\xi$. The gray area corresponds to the overdamped limit, where essentially $v_x^a$ depends solely on friction.
    }
    \label{fig:vel_core}
\end{figure}

In this paper, we present a detailed calculation of the flow around isolated $\pm 1/2$ defects and of the defect's self-propulsion velocity in an incompressible nematic film. We incorporate both viscous dissipation and frictional damping and examine the interplay between the two, as well as  the long range hydrodynamic effects arising from  incompressibility. We evaluate the $+1/2$ self-propulsive speed $|\fet v_+^a|$ as a function of the  the hydrodynamic dissipation length $\ell_d=\sqrt{\eta/\Gamma}$  which measures the competition between viscous dissipation and frictional damping. The result is summarized in Fig.~\ref{fig:vel_core}. When dissipation is controlled by friction ($\ell_d\ll \xi$),  one recovers the simple dimensional estimate $|\fet v^{a}_+|\sim|\alpha_0|/(\xi\Gamma)$. We show, however, that to obtain this result it is not sufficient to consider the far flow field which diverges near the defect, but one must resolve the full flow field near the defect core. On the other hand, when viscous stresses dominate, the defect propulsive speed depends on the order of limits. If $\Gamma=0$ from the outset, then a simple estimate yields $v_x^a\sim r$ due to the long-range nature of defect distortions. This limit, however, corresponds to a ``floating''  layer and does not describe experimental situations where the active nematic film is supported by a substrate~\cite{pismen2017viscous} or in contact with other fluids. It has been argued before that this unbounded growth should be cut off either by the system size or by the defect separation~\cite{giomi2014defect}. 
Our work shows that a finite friction cuts off the large scale divergence of the defect self-propulsion speed  at the scale $\ell_d$, with  $|\fet v_+^a|\sim\frac{|\alpha_0|}{\eta}\ell_d$  in the limit $\zeta=\ell_d/\xi\gg 1$ where viscous dissipation exceeds frictional drag and provides an analytical expression for the defect self-propulsion over all values of friction and viscosity.  
We  find that the structure of the flow field around a defect is also affected by the competition between viscosity and friction. At distances large compared to $\ell_d$, the flow velocity decays in the far-field as $\sim 1/r$, due to friction with the substrate ~\cite{pismen2017viscous}. At distances smaller than $\ell_d$, viscous dissipation dominates and  smooths out the velocity field near the defect core. Our work is relevant to defects in thin film of microtubule nematics on a substrates, as well as to dense cell layers.
%%%%%%%%%%% fig 1 %%%%%%%%%%%%%%%%%%%%%%
%%%%%%%%%%%%%%%%%%%%%%%%%%

In Sec.~\ref{Sec:hydro} we describe the hydrodynamic model. In Secs. \ref{Sec:Pos_Defect} and \ref{Sec:Neg_Defect}, we provide analytical derivations of closed expressions for the velocity and pressure fields induced by $\pm 1/2$ defects in an infinite system. One implication of the long-range interactions present in active nematics is that there are strong finite-size effects on the single defect flow field. This is discussed in Sec. \ref{Sec:numerics}, where we compare the analytical predictions with numerical integration of the Stokes equations in a disk of finite radius. Finally, the main results are discussed with concluding remarks in Sec. \ref{Sec:Conclusion}.
%%%%%%%%%%%%%%%%%%%%%%%%%%%%%%%%%%%%%%%%%%%%
\section{Hydrodynamic model}\label{Sec:hydro}
We consider a hydrodynamic model of an active nematic that couples flow velocity $\fet u(\fet r)$ to  the nematic order parameter $Q_{ij} = S (\hat n_i \hat n_j-\frac{1}{2} \delta_{ij})$, where $S$ quantifies the degree of order and $\mathbf{\hat n}(\mathbf{r})=\left(\cos\theta(\mathbf{r}),\sin\theta(\mathbf{r})\right)$ is the orientational director field with head-tail symmetry. In the simplest formulation, we consider that the $\textbf{Q}$-tensor is a minimizer of the de Gennes-Landau free energy ~\cite{marchetti2013hydrodynamics}
 \begin{align}
\mathcal F = \int d\fet r \left[\frac{K}{2}|\nabla Q|^2+\frac{g}{4} \left(1-\frac{1}{2}\textrm{Tr}\left(Q^2\right) \right)^2\right]\;,
\label{eq:F}
 \end{align}
with isotropic elastic constant $K>0$ and $g$ the strength of the local ordering potential. The uniform nematic ordered state  corresponds to $S_0^2=1$. The  flow field  satisfies a  Stokes equation that balances forces on a fluid element, given by~\cite{marchetti2013hydrodynamics}
\begin{align}
    \left(\Gamma-\eta \nabla^2 \right) \fet u =  \alpha_0 \nabla\cdot Q(\fet r) -\nabla p(\fet r), \qquad \nabla\cdot \fet u =0\;,
    \label{eq:Stockes_units}
\end{align}
where  $\Gamma$ is a friction coefficient per unit area,  $\eta$ is the dynamic viscosity, and $\alpha_0$ is the activity parameter, with dimensions of stress. For simplicity, we neglect the elastic stress as being of higher order in the gradients of $\mathbf{Q}$ compared to the active stress and a more important contribution for nematic textures with many defects. Here, we consider the flow field generated by an isolated $\pm 1/2$ defect embedded in an otherwise uniform nematic field. 

In two dimensions, the traceless  $\mathbf{Q}$-tensor has two independent components and can be represented equivalently as a complex scalar order parameter $\psi = Q_{xx}+iQ_{xy}$. The  configuration of a defect located at the origin can be written in terms of the $\psi$-field  as $\psi(\fet r) = S(r)e^{2i\theta(\fet r)}$, where $r\equiv |\fet r|$. The detailed form of core function $S(r)$ depends on the specific terms retained in the free energy, but it has the important generic asymptotic behaviors that $S(r) \rightarrow 1$ for $r\gg \xi$ and $S(r) \approx a r /\xi$ when $r\rightarrow 0$, where $\xi = \sqrt{K/g}$ is the coherence length which sets the scale of the defect core and $a$ is a numerical constant  $\mathcal O(1)$. Below we  set $a=1$, without loss of generality. The coherence length provides an ultraviolet cutoff to separate  inner core-solution from outer-core solution. On long distances, the nematic orientation is a potential field that has a branch cut starting at the origin where there is an isolated defect of charge $q= \pm 1/2$ and can be written as \cite{pismen1999vortices, angheluta2021role}  
\begin{align}
  \theta(\fet r) = q \arctan\left(\frac{y}{x}\right)+\theta_0,
\end{align}
where $\theta_0$ is the uniform background orientation. Without loss of generality, we set $\theta_0=0$.

We rescale the Stokes equation in units of the nematic relaxation time $\tau =\gamma/g $ (where $\gamma$ is the inverse of the rotational diffusivity) and the coherent length $\xi$, such that the dimensionless momentum equation takes the form 
\begin{equation}
        \left(1- \zeta^{2}\nabla^2 \right)\fet u =  \fet F^{\pm} -\nabla \tilde{p}(\fet r), \qquad \nabla\cdot \fet u =0,
    \label{eq:Stockes}
\end{equation}
where $\fet F^{\pm} = \alpha \nabla\cdot Q$ is the active force generated by a defect. 
The rescaled activity and pressure are  given by $\alpha =\alpha_0\gamma/(\Gamma K)$ and $\tilde{p}=p\gamma/(\Gamma K)$.  The dimensionless parameter  $\zeta =\sqrt{ \eta/(\Gamma\xi^2)} = \ell_d/\xi$ measures  the hydrodynamic dissipation length $\ell_d = \sqrt{\eta/\Gamma}$ in units of the coherent length, $\xi$. In the following, we will omit the tilde and all quantities are dimensionless unless otherwise stated.

The components of the $\mathbf{Q}$ tensor for an isolated $+1/2$ defect are given by $Q_{xx}(\fet r) = S(r) \frac{x}{r}$ and $ Q_{xy} = S(r) \frac{y}{r}$. The active force density then reduces to
\begin{align}
     \fet F^{+}(\fet r) = 
    \begin{cases}
    2\alpha \fet e_x , \quad r \rightarrow 0, \\
    \frac{\alpha}{r} \fet e_x  , \quad  r \gg 1.
    \end{cases}
\end{align}
Similarly,  for a negative defect $Q_{xx} = S(r) \frac{x}{r}$ and $Q_{xy} = -S(r) \frac{y}{r}$,  corresponding to an active force density given by 
\begin{align}
    \fet F^- (\fet r) = 
    \begin{cases}
    0 ,  \quad  r \rightarrow 0, \\
    -\alpha\frac{x^2-y^2}{r^{3/2}} \fet e_x + \alpha\frac{2xy}{r}\fet e_y,  \quad r \gg 1\;.
    \end{cases}
\end{align}
The  solutions for the flow velocity and pressure can be written in term  of the corresponding Green functions as
\begin{align}
    & \fet u (\fet r) = \frac{1}{2\pi\zeta^{2}} \int d\fet r' K_0( |\fet r-\fet r'|/\zeta) \left[\fet F^\pm(\fet r') -  \nabla' p(\fet r') \right] \equiv \fet u^{a}+\fet u^{p}, \label{eq:u_1}\\
    &p(\fet r) = \frac{1}{2\pi} \int d\fet r' \ln\left(|\fet r -\fet r'|\right) \nabla' \cdot \fet F^\pm (\fet r').\label{eq:pressure}
\end{align}
where $\fet u^{a}$ and $\fet u^{p}$ are the contributions to  the flow velocity induced by the active stress  and   pressure gradients, respectively. Note that the latter   also depends (indirectly) on activity. 
In the limit of no friction, Eqs. (\ref{eq:u_1}) and (\ref{eq:pressure}) reduce to  Eqs. (3.7) and (3.8) of Ref.~ \cite{giomi2014defect}.

%%%%%%%%%%%%%%%%%%%%%%%%%%%%%%%
\section{Positive nematic defect in an infinite system}\label{Sec:Pos_Defect}
\subsection{Defect self-propulsion}
The net active flow at the defect core acts as an advective velocity that propels the defect with a velocity $\fet v^{a}$, which in turn is controlled by both the active stress and  pressure gradients. Thus we write $\fet v^{a} = \fet u^{a}(0)+\fet u^{p}(0)$. The flow induced by the active stress at the origin is given by  Eq. (\ref{eq:u_1}) evaluated at $\fet r=0$. The $y$-component vanishes due to symmetry considerations, and the $x$-component is given by
\begin{align}
    &u_x^{a}(0)  = 2 \alpha \left[1 - \frac{1}{\zeta} K_1(\zeta^{-1} ) \right]
    +\frac{\pi\alpha}{2\zeta}   \left[1 -\frac{1}{\zeta}\left(  L_{-1}(\zeta^{-1}) K_0(\zeta^{-1}) + L_{0}(\zeta^{-1}) K_1(\zeta^{-1}) \right) \right]\;,
\end{align}
where $\zeta=\ell_d/\xi$, $K_n(x)$ are modified Bessel functions and $L_n(x)$ modified Struve function.

\begin{figure}[h]
    \centering
    \includegraphics[width =0.45\textwidth]{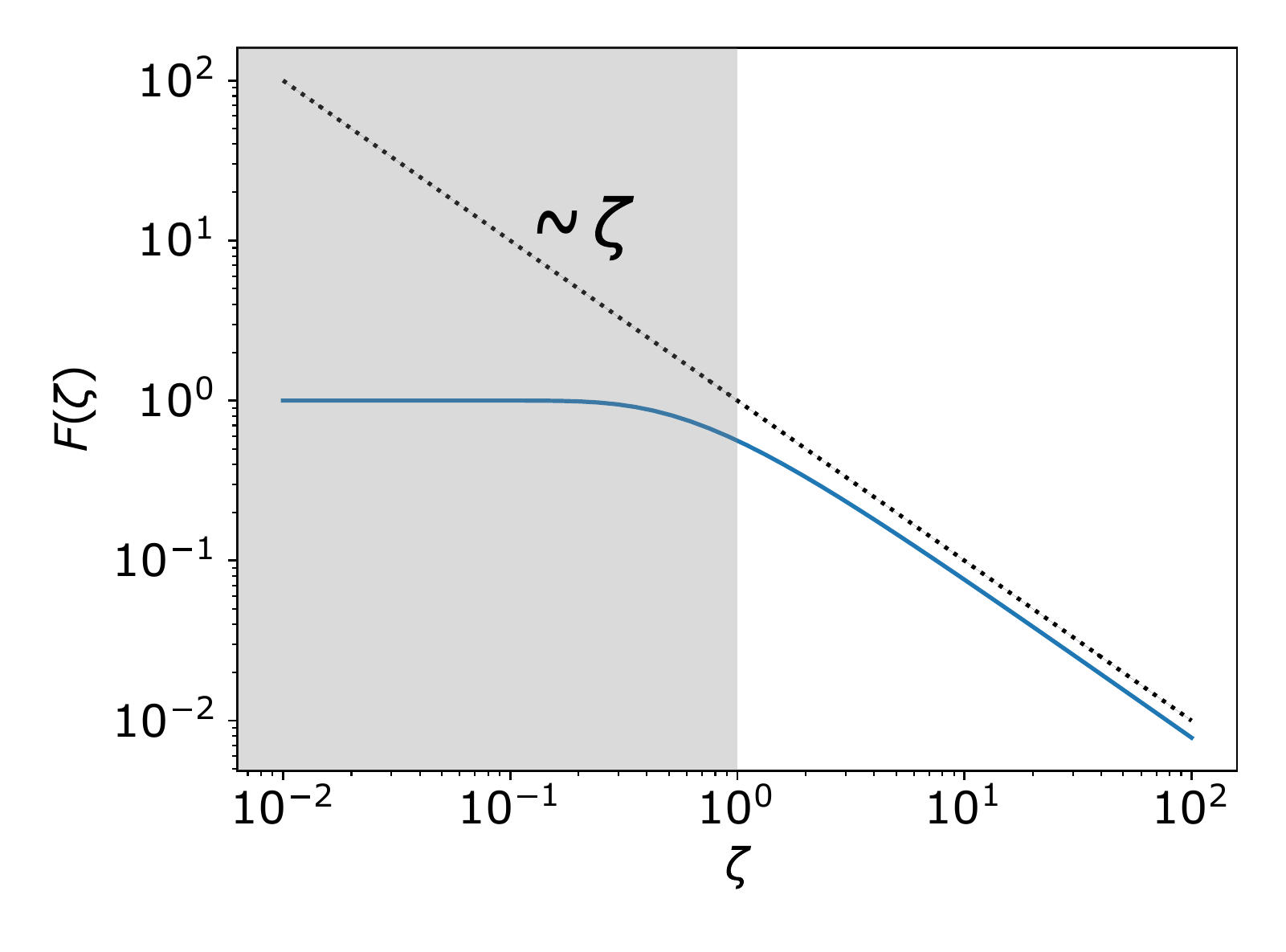}
    \caption{Scaling function $F(\zeta)$ as a function of $\zeta$. The gray region corresponds to the overdamped limit where $v_x^a$ depends only on friction. The doted black line is the asymptotic limit for $\zeta \gg 1$. }
    \label{fig:Scalefunction}
\end{figure}

The integral determining the pressure field given by Eq. (\ref{eq:pressure}) can be performed by a mapping to complex coordinates $(x',y')\rightarrow (w,\bar w)$, $(x,y)\rightarrow (z,\bar z)$ and then using the substitution to polar coordinates $w= r' \hat w$, $\hat w = e^{i\theta'}$. This yields 
\begin{align}
     p(\fet r) = &-\frac{\alpha }{2i\pi}\int_0^1 dr' r'\oint_\gamma d\hat w \left(\frac{1}{\hat w r'(\hat w -  zr^{'-1})} - \frac{1}{\bar z(  \hat w -r' \bar z^{-1})}\right) \nonumber\\
     & -  \frac{\alpha }{4i\pi}\int_1^\infty dr' \oint_\gamma d\hat w \left(\frac{1}{\hat w r'(\hat w -  z r^{'-1})} - \frac{1}{\bar z( \hat w -r' \bar z^{-1})}\right)
\end{align}
with $\gamma$ a contour of unit radius centered at origin. The pole at $\hat w=0$ is always inside the unit disk $|\hat w|<1$, whereas the poles at $\hat w = zr^{'-1}$ and $\hat w=r'\bar z^{-1}$ are inside the unit disk when $|z| < r'$ or $|z| > r'$, respectively. The contour integrals are then evaluated using the residue theorem.  Integrating over $r'$, we finally obtain
\begin{equation}
    p(\fet r) = 
    \begin{cases}
    \alpha  x , & r < 1, \\
    \frac{\alpha x}{r}   , & \text{if } r > 1.
    \end{cases}
\end{equation}
Consequently, the defect self-propulsion induced by pressure gradient has only  $x$-component which counteracts that induced by the active stress, and given by 
\begin{equation}
   u^{p}_x(0) = -\alpha \left(1 - \frac{1}{\zeta} K_1\left(\zeta^{-1} \right) \right) -   \frac{\pi \alpha}{4\zeta} \left[1 - \frac{1}{\zeta} [ L_{-1}(\zeta^{-1}) K_0(\zeta^{-1}) + L_{0}(\zeta^{-1}) K_1(\zeta^{-1}) ]\right]
   =-\frac{u^{a}_x(0)}{2}\;.
\end{equation}
Combining these results, we find that the  self-propulsion velocity of an isolated $+1/2$ defect oriented along the $x$ axis is $\mathbf{v}^a=v_x^a\hat{\mathbf{e}}$, where $v_x^a$ has the following scaling form
\begin{equation}
     v_{x}^{a} = \alpha  F(\zeta).
\end{equation}
where 
\begin{align}
     F(\zeta) =  \left(1 -\frac{1}{\zeta} K_1\left(\zeta^{-1}\right) \right) +\frac{\alpha  \pi}{4\zeta }\left[1 -\frac{1}{\zeta}[L_{-1}(\zeta^{-1}) K_0(\zeta^{-1}) + L_{0}(\zeta^{-1}) K_1(\zeta^{-1}) ]\right].
\end{align}
When $\zeta \gg 1$, we can simplify the expression by expanding in powers of $\zeta^{-1} $, and, to leading order, we obtain, 
\begin{equation}
 F(\zeta) \underset{\zeta\gg 1}{\approx} \frac{\pi }{4\zeta } +\frac{1 }{2\zeta^{2}}   (\gamma -1 -\ln(2\zeta) )
  -\frac{1}{4\zeta^{2}}   (2\gamma -1 - 2\ln(2\zeta))\;,  
  \label{eq:vel_at_center}
\end{equation}
where $\gamma \approx 0.577$ is the Euler-Mascheroni constant. Similarly, we also take the other limit $\zeta\ll 1$, where the scaling function approaches a constant value. The dependence of the scaling function $F$ on $\zeta$ is plotted in Fig.~\ref{fig:Scalefunction} and its asymptotic scaling at $\zeta \gg 1$ as $F \sim \zeta^{-1}$ is included as the dotted line. We can discuss the implications of there results better, when we use dimensional quantities and write the asymptotic behavior of the self-propulsion speed as 
\begin{equation} \label{eq:selfprop_speed}
    v_x^{a} \approx
    \begin{cases} \frac{\pi}{4}\frac{\alpha_0}{\Gamma \ell_d}= \frac{\pi}{4}\frac{\alpha_0 \ell_d}{\eta}, &  \zeta\gg 1\\
    \frac{\alpha_0 }{\Gamma \xi}, & \zeta\rightarrow 0
    \end{cases}
\end{equation}
 As anticipated from dimensional analysis, $v_x^a\sim \frac{\alpha_0}{\Gamma\xi}$, in the overdamped limit where dissipation is controlled only by frictional drag~\cite{shankar2018defect,shankar2019hydrodynamics, angheluta2021role}. In the underdamped limit, where the effect of drag is much smaller than viscous dissipation, hydrodynamic lenghscale becomes important in screening the divergence of the self-propulsion speed with system size, such that $v_x^a$ scales instead as $v_x^a\sim\alpha_0/\sqrt{\eta\Gamma}$. In this case, the self-driven motion of $+1/2$ defect is reduced by both friction and viscosity.
 
 %%%%%%%%%%% figure %%%%%%%%%%%%%%%

As discussed in the introduction, the presence of a finite drag always cuts off the large-scale divergence of the speed of a single defect obtained in a purely  viscous $2D$ layer at the dissipation length $\ell_d$.
When the flow equations for a thin nematic film of thickness $h$ on a substrate are derived via a lubrication approximation, the effective friction coefficient relates to the film thickness and the viscosity of the substrate bulk fluid (oil), and scales as $\Gamma\sim \bar\eta/h^2$~\cite{maitra2018nonequilibrium}. A more detailed calculation relevant to active microtubule suspensions confined between water and oil shows that the bulk viscosity plays an important role as an additional source of dissipation in the nematic layer affecting  the individual defect self-propulsion~\cite{guillamat2016probing}, as well as the vortex statistics in the active turbulence regime ~\cite{martinez2021scaling}. Notice that Ref. \cite{guillamat2016probing} shows that the $+1/2$ defect speed decays algebraically with the bulk oil viscosity (that controls the drag) in the regime where the flow dissipation comes from the viscous dissipation in the nematic layer, consistent  with our formulation.  When the flow dissipation is dominated by the oil bulk viscosity, there is, however, a logarithmic decay with increasing oil viscosity and, indirectly, drag.

%%%%%%%%%%%%%%%%%%%%%%%%%%%%%%%%%%%%%%%%%%%%%%
\subsection{Flow field away from the defect}
Outside the core, we treat the defect as a point source. From symmetry considerations, the flow velocity due to $\sigma^{a}$ is again non-zero only along the $x$-direction and it given by 
\begin{equation}\label{eq:ux_a}
    u_x^{a} (\fet r) =
    \frac{\alpha}{2\pi \zeta^{2}} \int \frac{d\fet r'}{r'} K_0(|\fet r'- \fet r|/\zeta ) \;.
\end{equation}

The flow velocity associated with pressure gradients is finite also in the $y$-direction and it is given by
\begin{equation}
    u^{p}_i(\fet r) = -\frac{\alpha}{2\pi \zeta^{2}} \int d\fet r' K_0( |\fet r' -\fet r|/\zeta ) \left( \frac{\delta_{ix}}{r'} -\frac{ x' r'_i}{r'^3} \right)\;.
    \label{eq:up}
\end{equation}
The term proportional to the $\delta$ function in Eq.~\ref{eq:up} cancels $\fet u^{a}_x$ from Eq. \ref{eq:ux_a}, such that the total active fluid flow is entirely determined by pressure flow, with
\begin{equation}
    u_i(\fet r) = \frac{\alpha}{2\pi\zeta^{2}} \int d\fet r' K_0(|\fet r' -\fet r|/\zeta )\frac{ x' r'_i}{r'^3}\;.
    \label{eq:positive_v_integral_form}
\end{equation}
To evaluate this integral, we use a complex representation  $u = u_x + i u_y$
and evaluate the resulting contour integrals as shown in appendix \ref{ap:Agony} where we express them in terms of complete elliptic integrals of first and second kind. We further use the power series representation of these elliptic integrals, which allows us to write the active fluid velocity as a series expansion in integrals over the zeroth order modified Bessel function, namely
\begin{align}
     u^+(r,\phi) =& \frac{\alpha}{2\zeta^{2} }
      \sum_{n=0}^{\infty}\left(
       1 - \frac{2n+1}{2n-1}  e^{2i\phi}  
    \right) \left( \frac{(2n -1)!!}{(2n)!!} \right)^2 
    \int_0^{r} dr'  K_0( r'/\zeta) \left(\frac{r'}{r} \right)^{2n+1 }
    \nonumber\\
    +&
    \frac{\alpha}{2 \zeta^{2}}\sum_{n=0}^\infty \left( 
    1 - \frac{n}{n+1} e^{2i \phi} 
    \right)  \left( \frac{(2n -1)!!}{(2n)!!} \right)^2  \int_{r}^{\infty} dr'  K_0(  r'/\zeta) \left(\frac{r}{r'} \right)^{2n }\;.
    \label{eq:VelPositiveIntegralExsp}
\end{align}
The  $K_0(x)$ integrals are computed in appendix \ref{ap:Truble}. After some mathematical manipulations the  velocity reduces to 
\begin{equation}
    u^+(r,\phi) = \frac{\alpha}{4\zeta} \left[ \pi \left(I_0(r/\zeta)-I_2(r/\zeta) e^{2i\phi}\right) + \sum_{k,n=0}^\infty \left( \kappa^+_1(n,k) +\kappa^+_2(n,k)  e^{2i\phi}\right)\frac{1}{ (k!)^2} \left(\frac{r}{2\zeta}\right)^{2k+1} \right].
    \label{eq:velocityfield_positive_defect}
\end{equation}
with
\begin{align}
    \kappa^+_1(n,k) &= \left( \frac{(2n -1)!!}{(2n)!!} \right)^2 \left( \frac{-(4n+1)(4k+3)}{(n+k+1)^2(2n-1-2k)^2}  \right)\;, \\
     \kappa^+_2(n,k) &= \left( \frac{(2n -1)!!}{(2n)!!} \right)^2  \frac{[(2n-1)(4k(n+1) +1+n) -4k^2] (4n+1)}{(n+1+k)^2(2n-1-2k)^2(n+1)(2n-1)}\;.
\end{align}
The corresponding vorticity  is given by 
\begin{align}
   \omega^+(r,\phi) = -\frac{\alpha}{8 \zeta^{2}}\sin(\phi) \left(4\pi I_1( r/\zeta) + \sum_{n,k} [( 2k+1) \kappa^+_1(n,k) -(2k +3)\kappa^+_2(n,k)]\frac{1}{(k!)^2 } \left(\frac{r}{2 \zeta}\right)^{2k} \right). 
   \label{eq:vorticityfield_positive_defect}
\end{align}
Both velocity and vorticity are shown in Fig.~\ref{fig:Positive_velocity_low_damping}. 
%%%%%%%%%%%%%%%%%%%%%%%%%%%%%%%%%%%%%%%%%%%
\subsubsection{Asymptotic far-field flow:}
The  flow field greatly simplifies in the far-field  $ r/\zeta \gg 1$, corresponding to distances much larger than the hydrodynamic dissipation length. Then, the second term in Eq.~(\ref{eq:VelPositiveIntegralExsp}) vanishes due to the exponential decay of the Bessel function. In the first integral, we can replace the upper limit $r$ with $\infty$ and perform it analytically with the result given as
\begin{align}
     u^+(r,\phi) = \frac{\alpha}{2r} \left(e^{2i\phi} +1  + \left(\frac{\zeta}{r}\right)^2 (1-3e^{2i\phi}) \right)\;,
    \label{eq:Vel_Asymptotic_Positive}
\end{align}
where we have kept the two first terms in the expansion.
%%%%%%%%%%% fig 2 %%%%%%%%%%%%
\begin{figure}[ht]
    \centering
    \includegraphics[width = 0.45\textwidth]{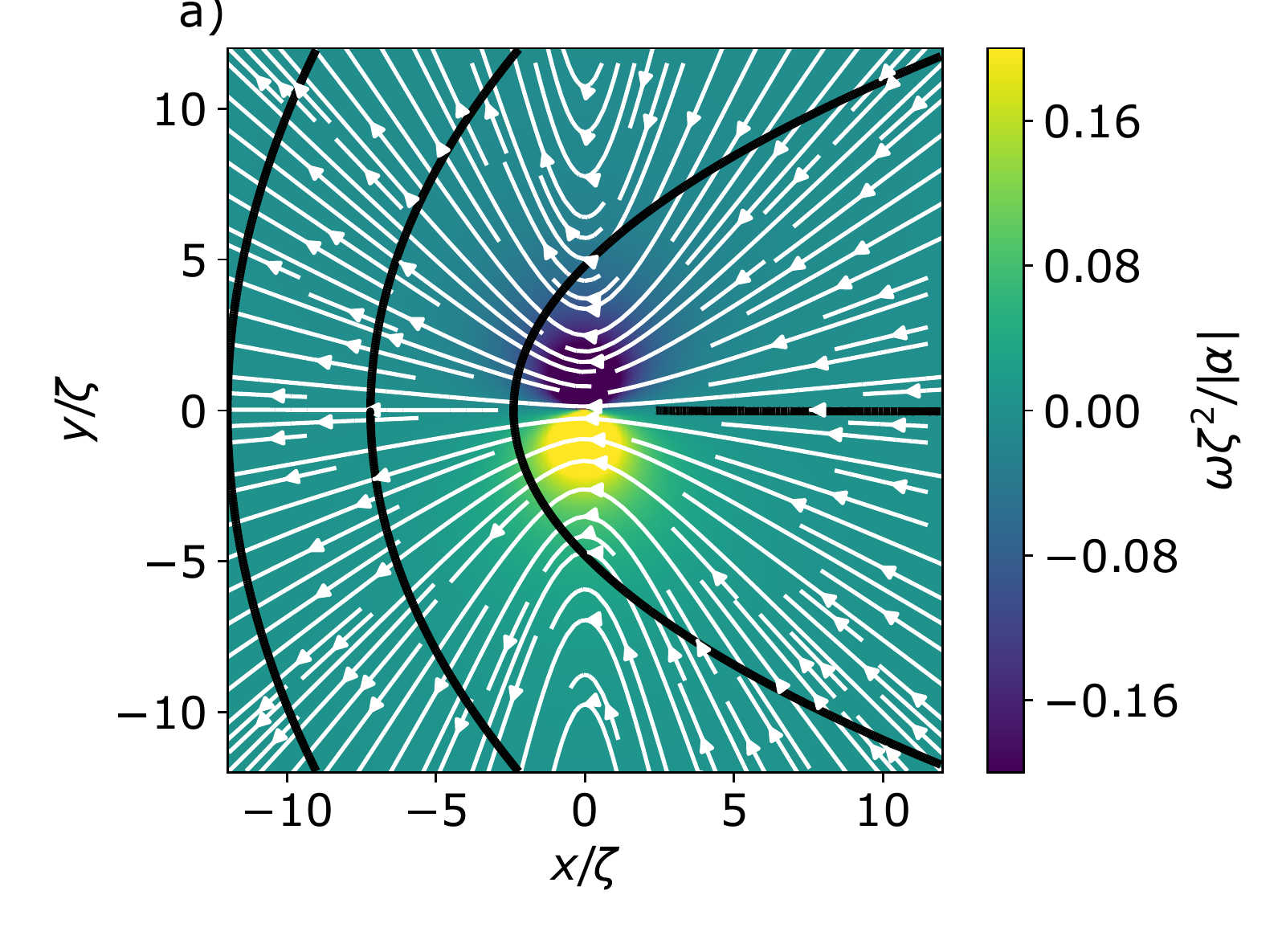}
    \includegraphics[width = 0.45\textwidth]{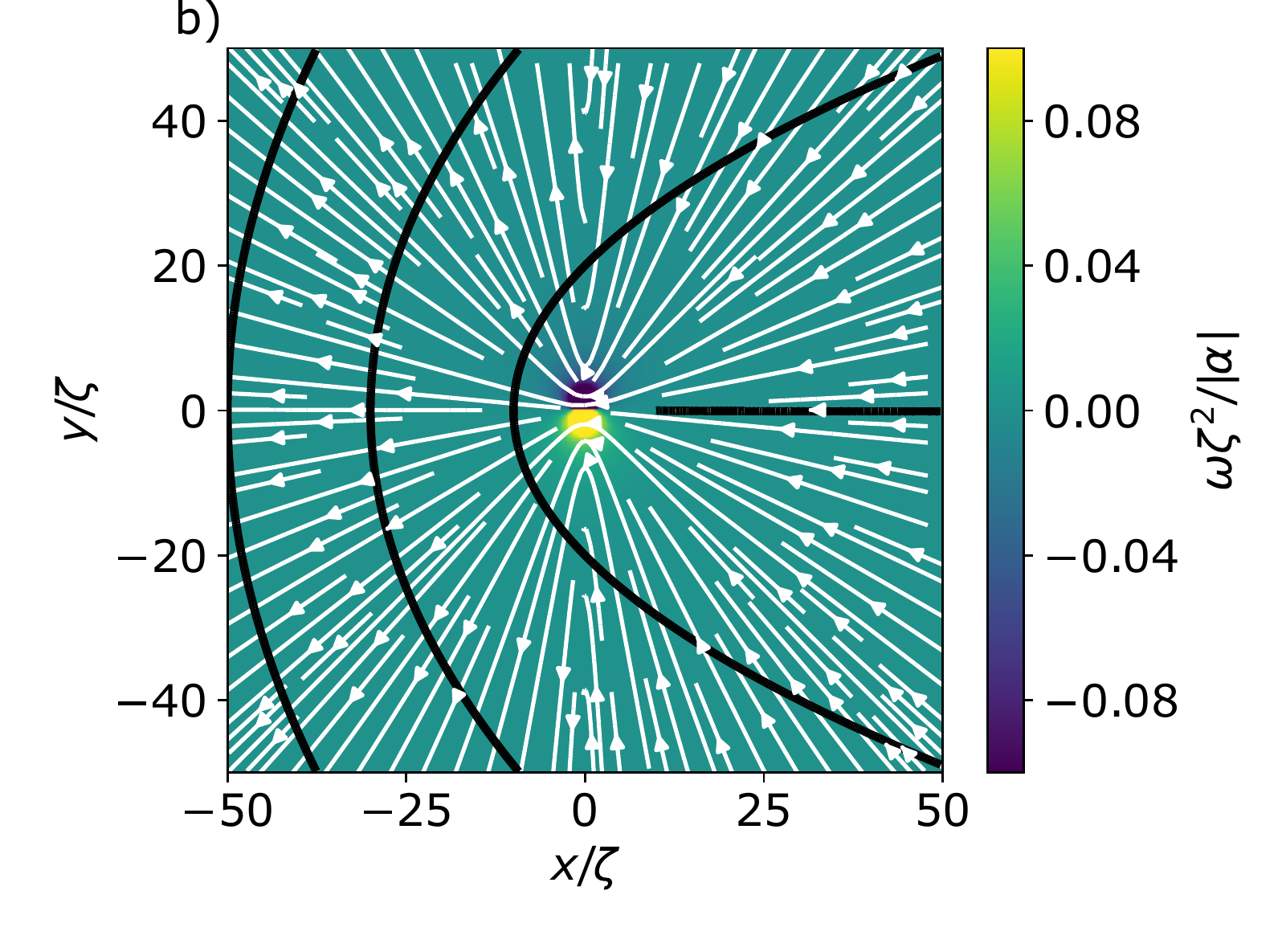}
     \includegraphics[width =0.45\textwidth]{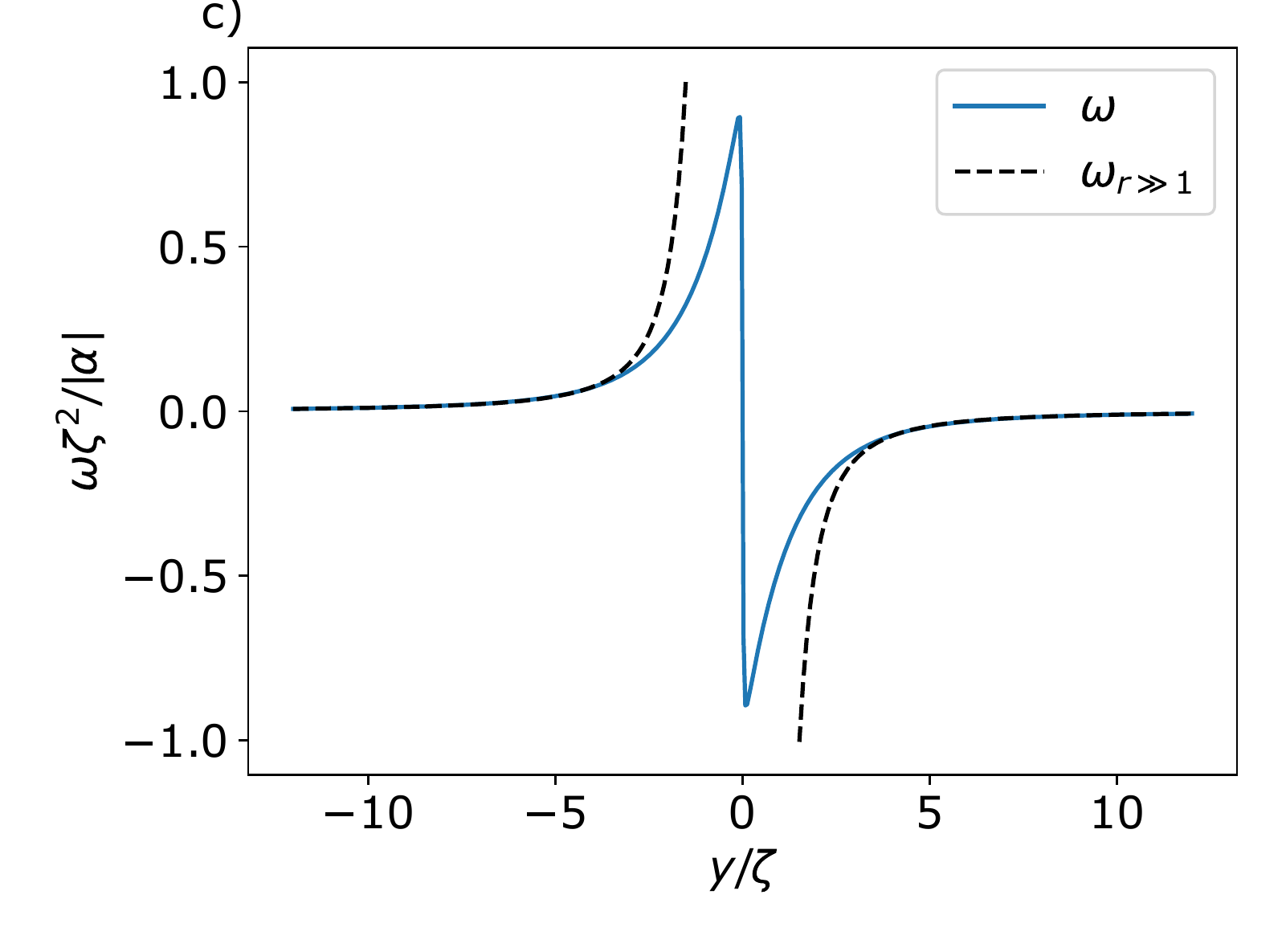}
     \includegraphics[width = 0.45\textwidth]{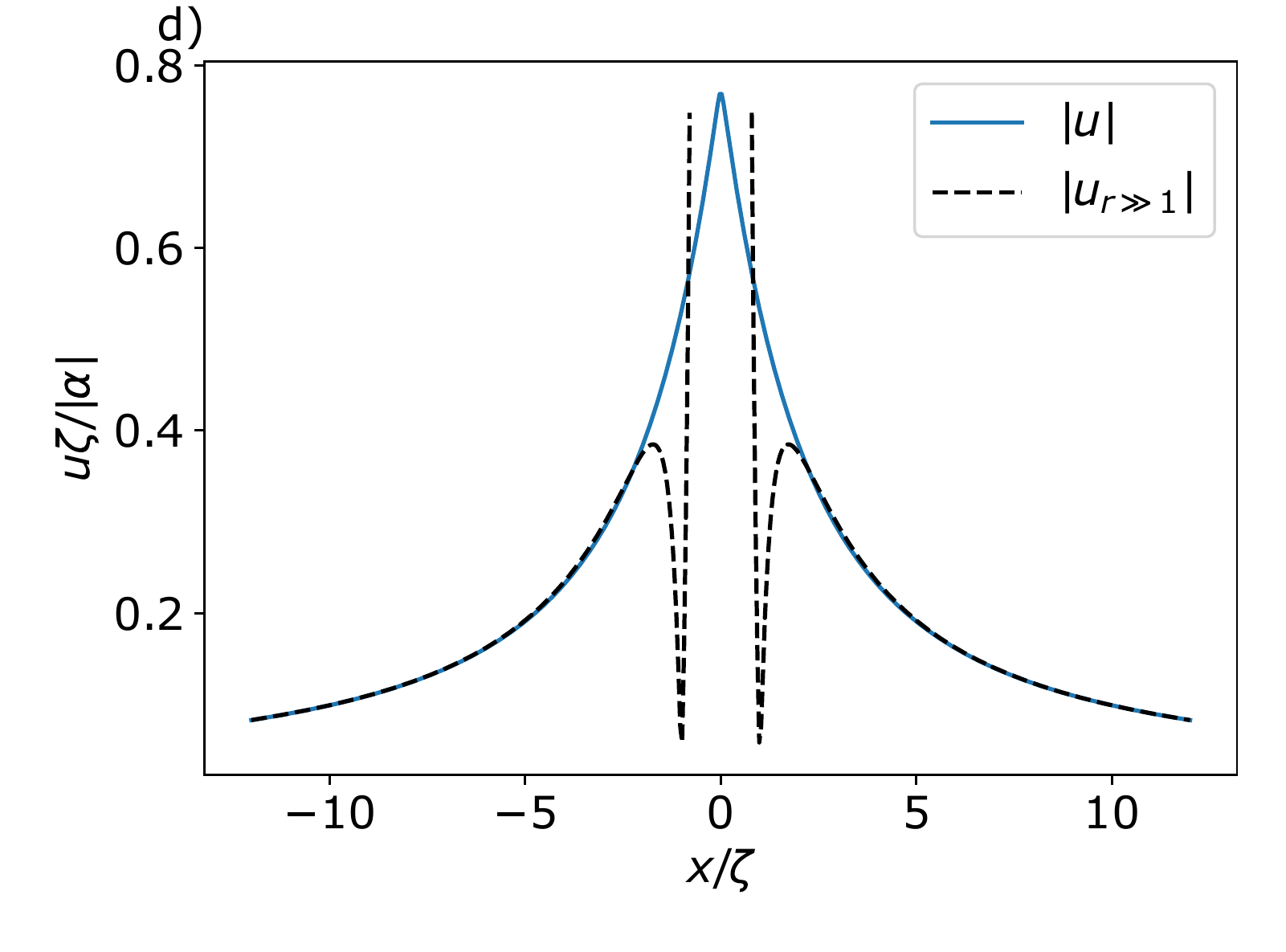}
    \caption{Flow streamlines (white arrow) around a $+1/2$ defect for $\alpha<0$ obtained from (a) full solution and (b) asymptotic one.
    The nematic director field is shown in black lines and the background colormap denotes vorticity. To show the structure of the near-field, the vorticity scale is saturated at $\pm 0.2$ in a) and at $\pm 0.1$ in b). c) Cross section of vorticity obtained from the exact solution (solid blue line) and the asymptotic limit (dotted black line) at $x=0$ as a function of $y$. d) Cross section of the velocity obtained from the exact solution (solid blue line) and the asymptotic limit (dotted black line) at $y=0$ as a function of $x$.}
    \label{fig:Positive_velocity_low_damping}
\end{figure}
%%%%%%%%%%%%%%%%%%%%%%%%%%%%%%%%%%%
 The slow $1/r$-decay term in Eq.~\ref{eq:Vel_Asymptotic_Positive} is independent of viscosity $\eta$ and identical to the one derived in Ref. \cite{angheluta2021role} in the friction-dominated regime. Corrections due to viscosity give rise to faster $1/r^3$ decay. The corresponding far-field vorticity is 
\begin{equation}
    \omega^+(r,\phi) =  \frac{\alpha}{r^2}\sin{\phi} \left(1  +3\left(\frac{\zeta}{r}\right)^2 \right)\;.
\end{equation}
The far-field solutions are singular at the origin, which is not the case for the full series solution that resolves the near core field. This is demonstrated visually in Fig. \ref{fig:Positive_velocity_low_damping} (c-d) where we plot cross-sections of the velocity and vorticity profiles for both the full solution and the far-field solution.
The form of the expressions makes it natural to scale the position, velocity and vorticity with $\zeta$, $\zeta/|\alpha|$ and $\zeta^2/|\alpha|$ respectively.
The only free parameter is then the sign of $\alpha$.
Panels (a) and (b) show the flow streamlines and the vorticity field in the background for the full and the far-field solutions,  respectively, for an extensile system ($\alpha<0$). The velocity magnitude is highest near the defect core and  decays as a power law following the far-field asymptote. The velocity streamlines point towards the defect in the right half-plane, and away from the defect in the left half-plane. For positive $\alpha$, the flow direction is reversed. In an infinite system, the flow streamlines around an isolated defect are not closed. On the other hand, as discussed later, in bounded domains, the  system size controls the size of the eddies formed around the defect. For more realistic configurations with many defects, the system size is typically replaced by the mean defect separation. It may be that other intrinsic length scales controlled by elastic stresses are also important in stabilizing finite-size vortices. These effects are left for future   investigation.

%%%%%%%%%%%%%%%%%%%%%%%%%%%%%%%%%%%%%%%%%%%%%%%%%%%%%%%%%%
\section{Negative nematic defect in an infinite system} \label{Sec:Neg_Defect}

By similar calculations as in Sec. \ref{Sec:Pos_Defect}, we find that the velocity induced by the active stress at the position of the negative defect vanishes as expected from symmetry consideration. After performing the integral in the complex plane and subsequently integrating over the integrand with the Bessel function, we determine the pressure field induced by the $-1/2$ defect vanishes inside the defect core and non-zero outside given by 
\begin{align}
    p(\fet r) = 
    \begin{cases}
    0, \quad r<1,\\
    -\alpha \frac{x^3 - 3xy^2}{3r^3}, \quad r>1.
    \end{cases}
\end{align}
and its gradient vanishes at the origin, hence no advective pressure-flow of the negative defect. Thus, an isolated $-1/2$ defect is stationary in a uniform nematic field, regardless of activity.   

%%%%%%%%%%%%%%%%%%%%%%%%%
\subsection{Flow field away from the defect:}
The flow field induced by the $-1/2$ defect can also be expressed analytically as a series expansion of the elliptic integrals as detailed in appendix \ref{ap:dreams}, with the resulting expression of the velocity field in the complex representation $u^-= u^-_x+i u^-_y$ given as 
\begin{align}
    u^-(r,\phi) = &-\frac{\alpha}{2\zeta^{2} } \sum_{n=0}^{\infty} \left( \frac{(2n -1)!!}{(2n)!!} \right)^2  \frac{2n+1}{2n-1} \left[ e^{4i\phi}  \frac{2n+3}{2n-3}
    -e^{-2i\phi} \right] 
    \int_0^{r}dr' K_0( r'/\zeta) \left(\frac{r'}{r}\right)^{2n+1} \nonumber
    \\
    &-\frac{\alpha }{2\zeta^{2}}\sum_{n=0}^{\infty}\left(\frac{(2n-1)!!}{(2n)!!}\right)^2 \frac{n}{n+1}  \left[ e^{4i\phi}  \frac{n-1}{n+2} - e^{-2i\phi} \right]  \int_{r}^{\infty}dr' K_0( r'/\zeta) \left(\frac{r}{r'} \right)^{2n}.
    \label{eq:Integral_eq_for_hydrodynamic_neg_defect}
\end{align}
The integrals over the Bessel functions are evaluated in appendix \ref{ap:Truble}, and the final expression is then given as  
\begin{align}
       u^-(r,\phi) = \frac{\alpha  }{8\zeta}
       \left(2\pi \left[I_2( r/\zeta)e^{-2i\phi} - I_4( r/\zeta)e^{4i\phi}\right] 
       +  \sum_{k,n}
       \left[\kappa^-_1(n,k) e^{-2i\phi} +  \kappa^-_2(n,k)e^{4i\phi}
       \right] \frac{2}{(k!)^2} \left(\frac{r}{2\zeta} \right)^{2k+1} \right)
       \label{eq:velocityfield_negative_defect}
\end{align}
with the coefficients 
\begin{align}
       &\kappa^-_1(n,k) = \left( \frac{(2n-1)!!}{(2n)!!} \right)^2 \frac{(4n+1)[4k^2 -(2n-1)(4k+1)(n+1) ]}{(2n-1)(n+1)(n+1+k)^2 (2n-1-2k)^2},\\
      &\kappa^-_2(n,k) = \left( \frac{(2n-1)!!}{(2n)!!} \right)^2 \Bigg[\frac{4n(n-1)}{(n+1)(n+2)(2n-1-2k)^2} 
       - \frac{(2n+1)(2n+3) }{(2n-1)(2n-3)(n+k+1)^2} \Bigg].
\end{align}
The corresponding vorticity field as function of the polar coordinates follows as, 
\begin{equation}
    \omega^-(r,\phi) = -\frac{\alpha }{8\zeta^{2}} \sin(3\phi) \left( 4\pi I_3( r/\zeta) + \sum_{k,n}\left[(2k-1)\kappa^-_1(n,k) - (2k+5)\kappa^-_2(n,k)\right] \frac{1}{(k!)^2 } \left (\frac{r}{2\zeta}\right)^{2k} \right)
    \label{eq:vorticityfield_negative_defect}
\end{equation}

%%%%%%%%%% fig 3 %%%%%%%%%%%%%%%%%%%%%%%
\begin{figure}[ht]
    \centering
        \includegraphics[width=0.45\textwidth]{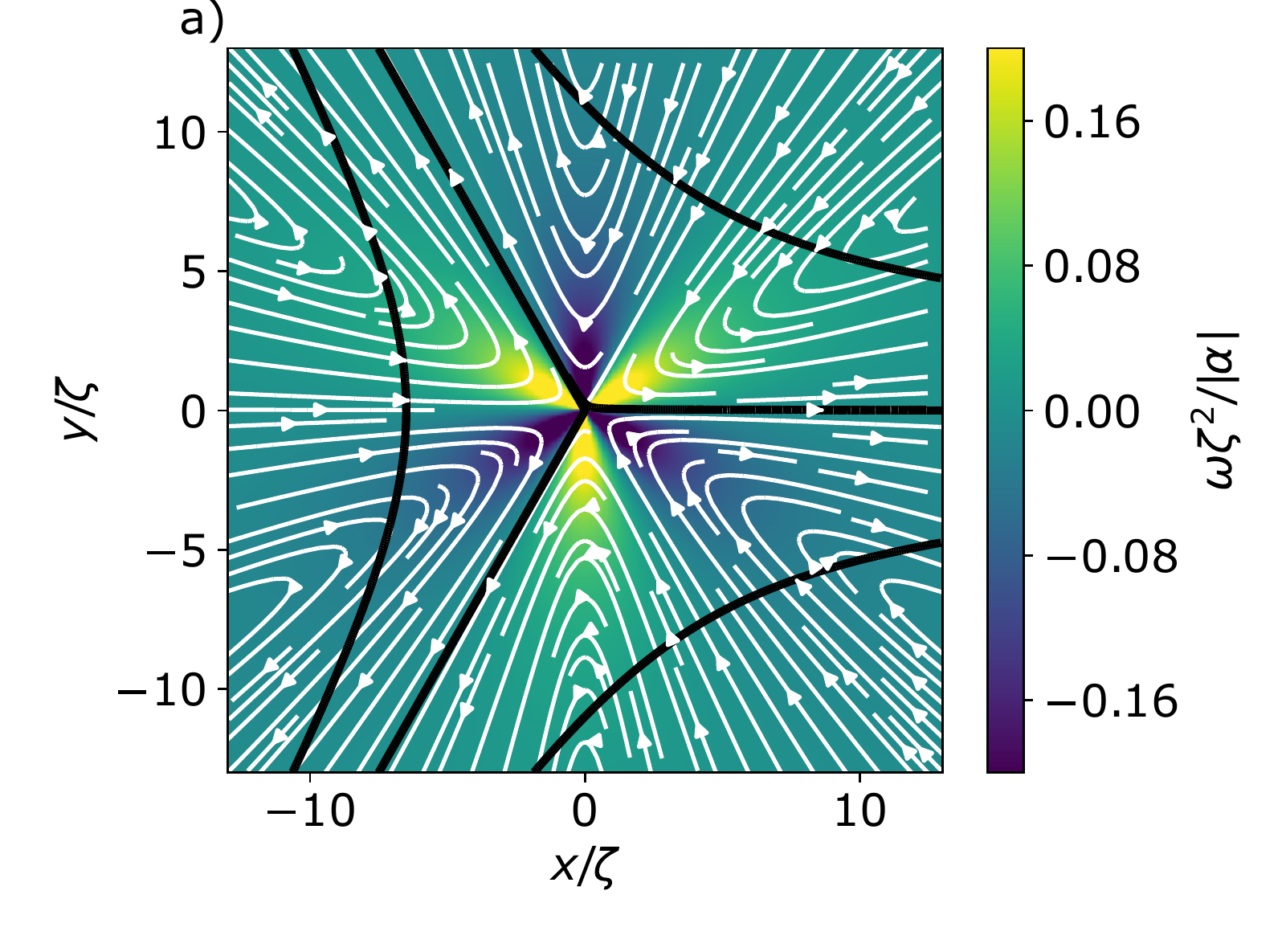}
    \includegraphics[width = 0.45\textwidth]{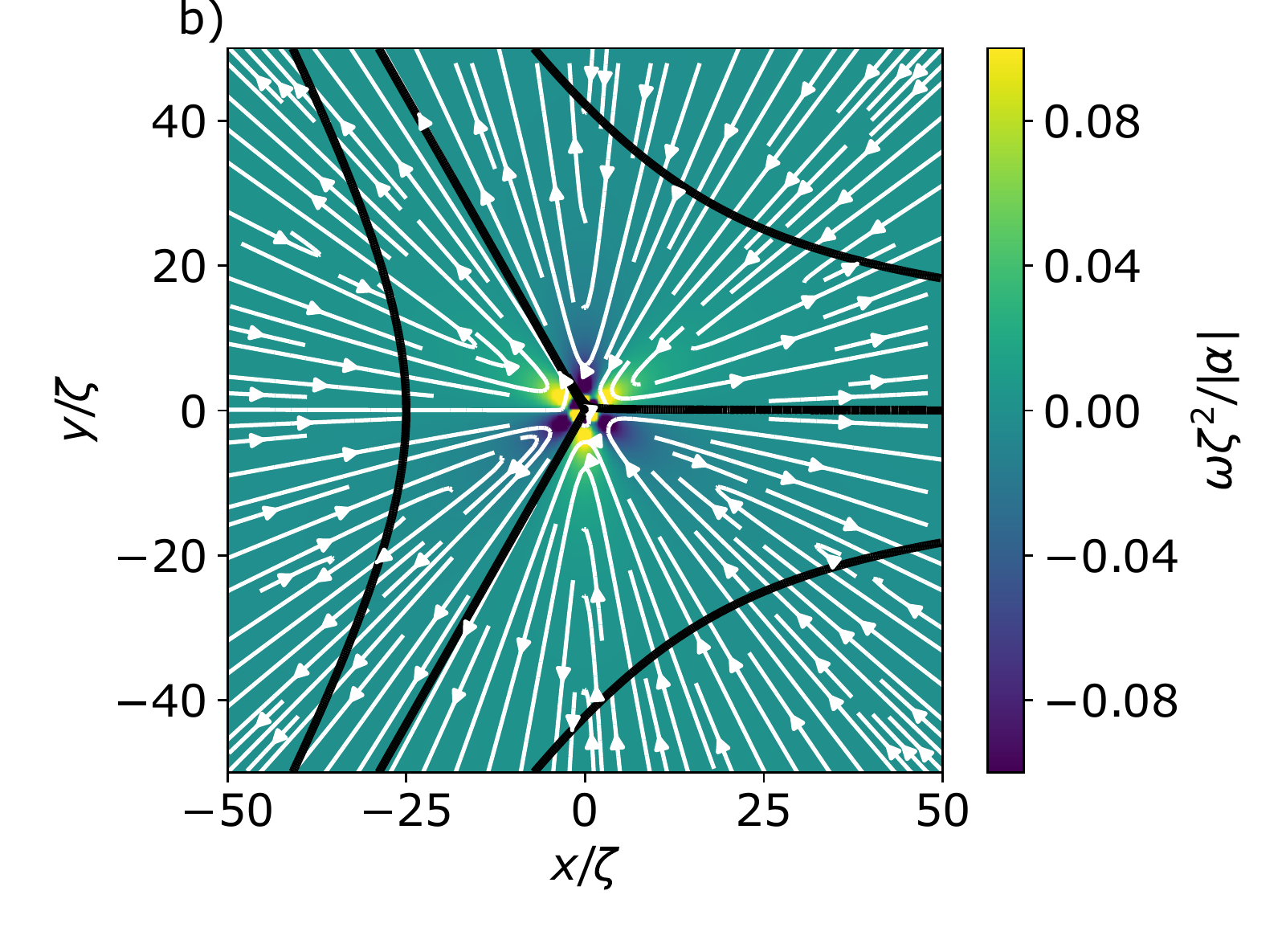}
    \includegraphics[width = 0.45\textwidth]{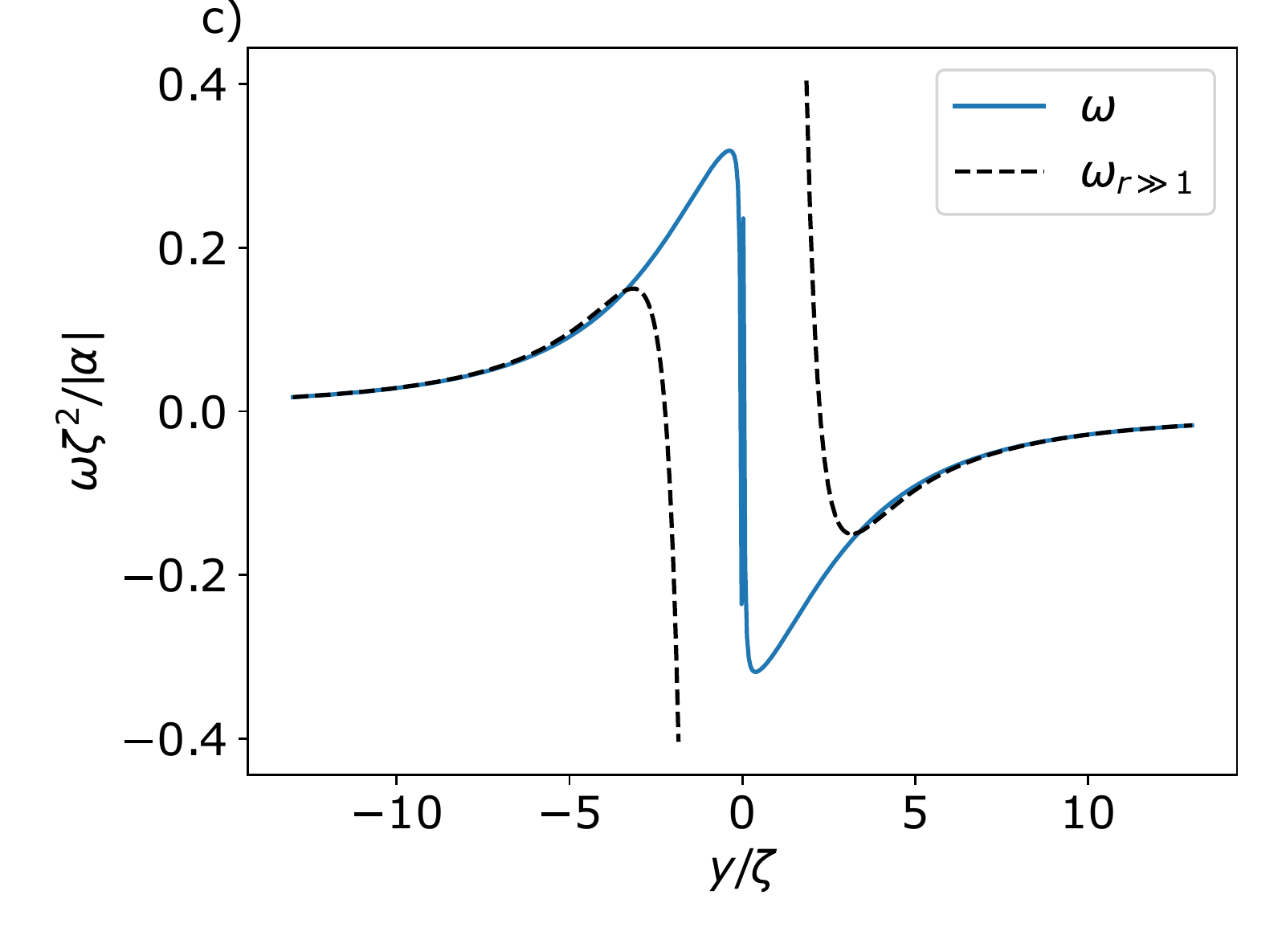}
     \includegraphics[width = 0.45\textwidth]{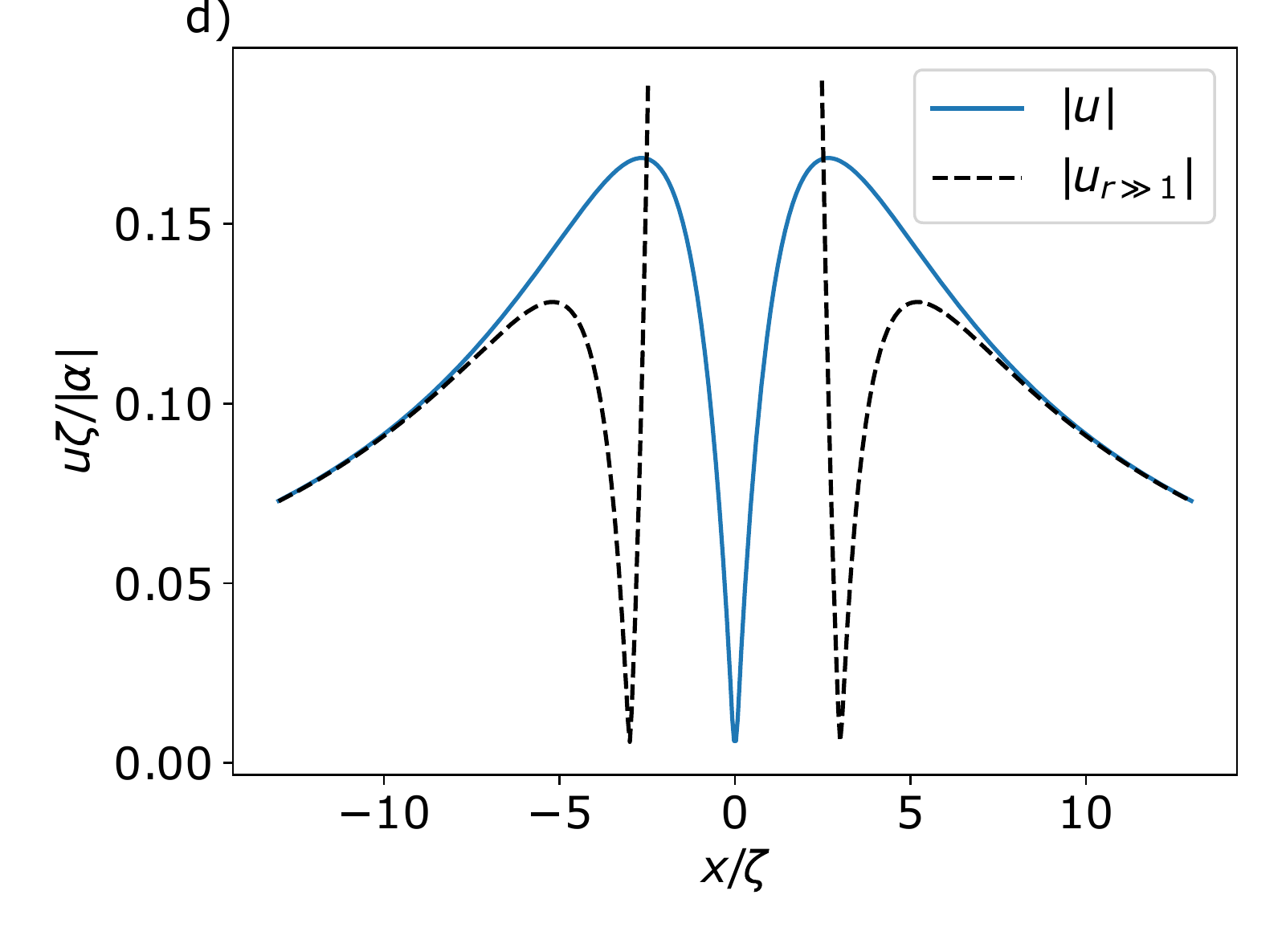}
    \caption{Flow streamlines (white arrow) around a $-1/2$ defect for $\alpha<0$ obtained from (a) full solution and (b) asymptotic one.
    The nematic director field is shown in black lines and the background colormap denotes vorticity. To show the structure of the near-field, the vorticity scale is saturated at $\pm 0.2$ in a) and at $\pm 0.1$ in b). c) Cross section of vorticity obtained from the exact solution (solid blue line) and the asymptotic limit (dotted black line) at $x=0$ as a function of $y$. Notice that the vortices changes sign rapidly at origin due to its multivalued phase (see Eq.~\ref{eq:vorticityfield_negative_defect}). d) Cross section of the velocity obtained from the exact solution (solid blue line) and the asymptotic limit (dotted black line) at $y=0$ as a function of $x$. }
    \label{fig:neg-high}
\end{figure}
%%%%%%%%%%%%%%%%%%%%%%%%%%%%%%%%%%%%%%
\subsection{Asymptotic far-field flow:}
As with the $+1/2$ defect, the far-field asymptotic flow is dominated by the leading order terms in the expansion, which can also be computed directly from Eq.~(\ref{eq:Integral_eq_for_hydrodynamic_neg_defect}) in the limit of $ r/\zeta \rightarrow\infty$. The result of this calculation is that 
\begin{align}
    u^-(r,\phi) \underset{ r/\zeta \gg 1}{\approx} 
     = \frac{\alpha}{2r} \left[
         \left(\frac{\zeta}{ r}\right)^2 \left(15 e^{4i\phi} + 3 e^{-2i\phi} \right) -\left( e^{4i\phi} +e^{-2i\phi} \right) \right]. 
\end{align}
As for eq. (\ref{eq:Vel_Asymptotic_Positive}) the $1/r$ term here was also obtained in Ref. \cite{angheluta2021role}. 
The vorticity related to this velocity is
\begin{equation}
     \omega^-(r,\phi) =  \frac{3\alpha\sin(3\phi)}{r^2}\left(5\left(\frac{\zeta}{ r}\right)^2-1\right).
 \end{equation}
In this asymptotic approximation, the flow field is singular at the origin. This singularity is however lifted by the higher order terms in the series expansions, so that the exact flow is smooth everywhere. Fig. \ref{fig:neg-high} shows the flow streamlines with the vorticity field as the colormap for the asymptotic (in panel a) and the exact solutions (in panel b),
with the values scaled in the same way as for fig. \ref{fig:Positive_velocity_low_damping}.
Cross-sections of the vorticity and velocity at $y=0$ are plotted in panels (c)-(d) showing the singular behavior of the asymptotic approximation at the origin, while it captures very well the far-field behavior. The plots correspond to an extensile system with $\alpha<0$. 
The $\sin(3\phi)$ factor in the vorticity divides the plane in six regions where the sign of the vorticity is altered and making it multi-valued at the origin. The size of the velocity is zero at origin as we discussed above. It increases a bit outside before it starts to decay with increasing $r$ following the far-field asymptotic behavior. 

As for the $+1/2$ defect, the flow streamlines never closed in an infinite system, thus there are no finite size vortices. In the next section, we discuss how the picture changes once the defect is placed in bounded domain.

%%%%%%%%%%%%%%%% figure 4 %%%%%%%%%%%%%
\begin{figure}[ht]
    \centering
   \includegraphics[width=1\textwidth, trim = 0.0cm 13.0cm 0.0cm 0.0cm, clip = true]{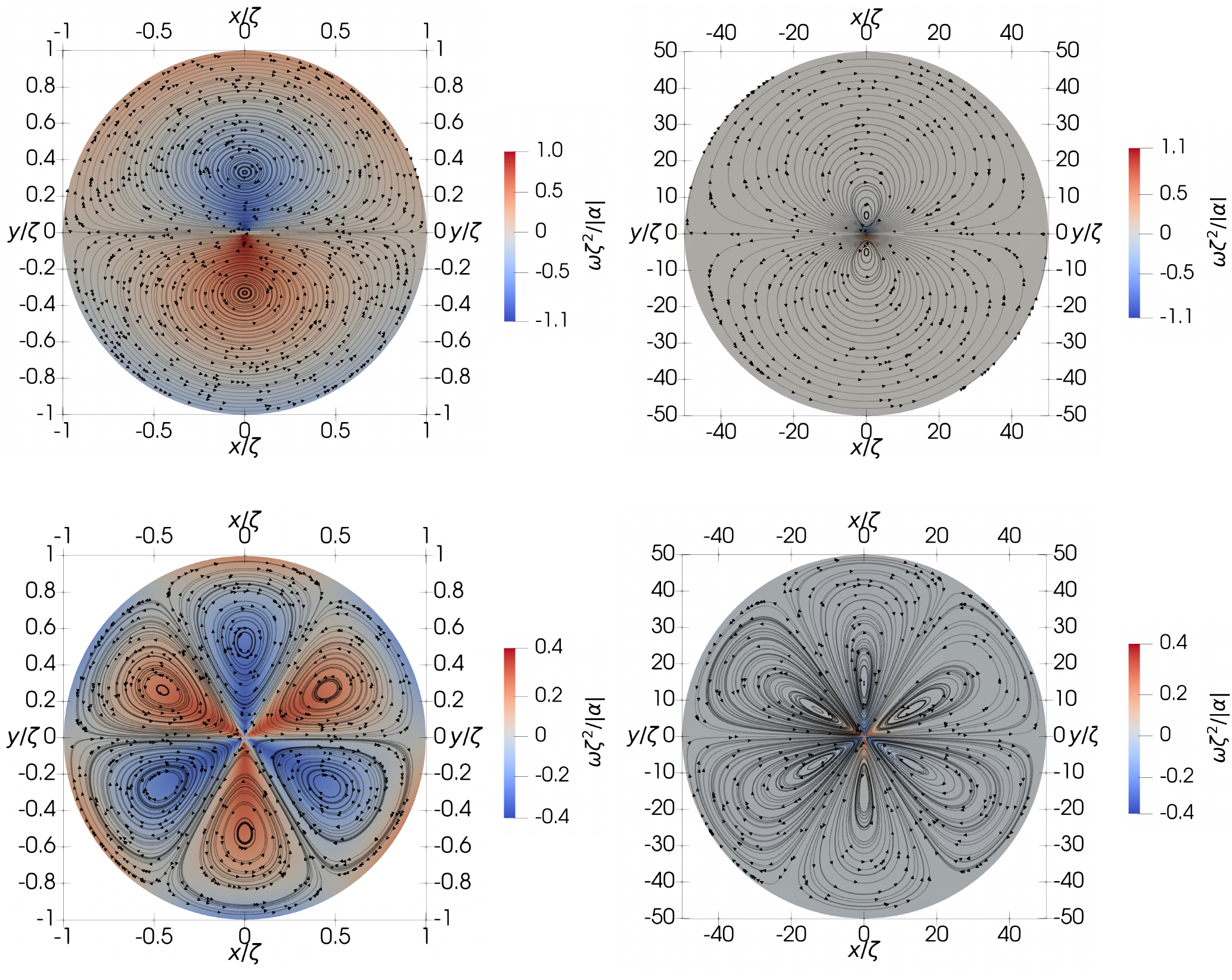}
    \caption{Flow streamlines (black lines) with vorticity as a colormap background generated by a $+1/2$ (a,b) and a $-1/2$ (c,d) defect  in disks of different radii $R$ for an extensile system ($\alpha<0$).  (a) and (c) are for $R= 1$  and (b) and (d) $R= 50$.}
    \label{fig:Numerics_stream_plots}
\end{figure}
%%%%%%%%%%%%%%%%%%%%%%%%

%%%%%%%%%%%%%%%%%%%%%%%%%%%%%%%%%%%%%%%%%%%%%%%%%%%%%
\section{Isolated defect in a bounded active nematic} \label{Sec:numerics}
The problem of finding the flow field around defects in a bounded domain is  challenging to solve analytically. Thus, we resort to numerical solutions of the Stokes flow given by Eq. ~(\ref{eq:Stockes}) in a disk of radius $R$ using finite element methods and homogeneous boundary conditions (zero velocity). In addition, we use the simplification that a single defect is imprinted in an uniform nematic field, while the changes in the nematic orientation induced by confinement are ignored~\cite{giomi2014defect}. 

The Stokes flow Eq.~\ref{eq:Stockes} is solved with FEniCS using Taylor-Hood elements, which are quadratic for the velocity and linear for the pressure and vorticity \cite{alnaes2015fenics,LoggMardalEtAl2012}. 

Fig.~\ref{fig:Numerics_stream_plots} shows the flow streamlines induced by a single $+ 1/2$ (a,b) and $- 1/2$ (c,d) defect in a disk of radius $R$ for an extensile system for $\eta\ne 0$ and $\Gamma\ne 0$. The left and right columns correspond to $R=1$ and $R=50$, respectively. In a  bounded system, the vortical flows around each defect span the system size, as also reported in Ref.~\cite{giomi2014defect} for $\Gamma=0$. However, due to friction with the substrate, the flow decays on length scales larger than $\ell_d$. This is evident by comparing the values in the far-field of vorticity in left and right columns from Fig.~\ref{fig:Numerics_stream_plots}, corresponding to $R=1$ (in units of $\ell_d$) in (a,c) and $R=50$ (in units of $\ell_d$) in (b,d). 
We notice that the center of a vortex is not fixed at the maximum of the vorticity. This is due to the fact that the $\pm 1/2$ defects generate shear flows which localise shear vorticity next to the defect cores. However, unlike curvature vorticity in rotating flows which peaks at the vortex core, shear vorticity is not necessarily an indication of the presence of vortices or their location. In fact, with increasing $R$, the flow gradients near the defect cores become sharper, the streamlines near the cores are "stretched" in the radial direction, and the eyes of vortices move further from the origin. In the  limit  $R \rightarrow \infty$, we expect vortices to get stretched out so that flow streamlines close at infinity, and we recover the analytic flow profiles. 

\begin{figure}
    \centering
    \includegraphics[width = 0.48\textwidth]{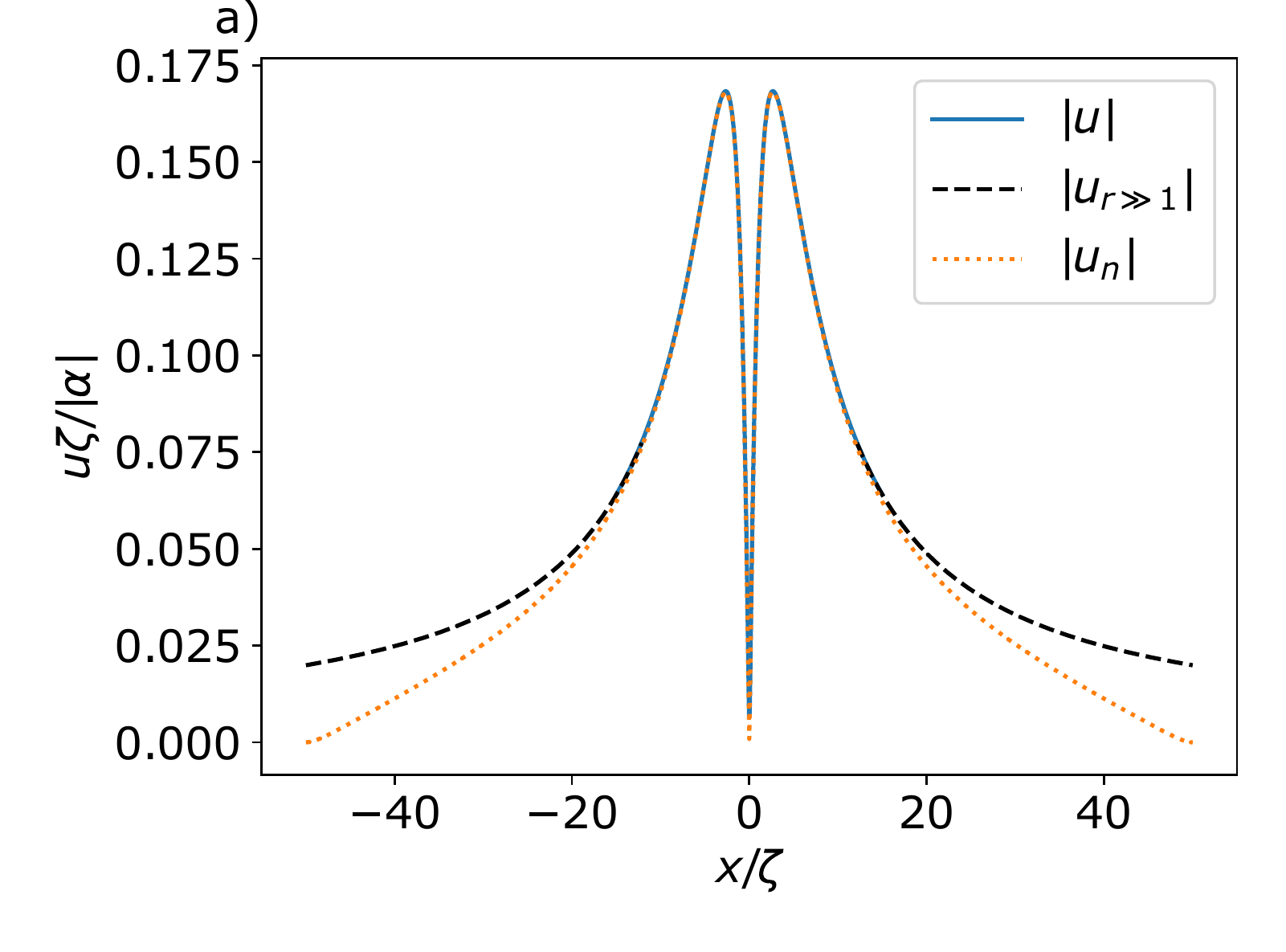}
    \includegraphics[width = 0.48\textwidth]{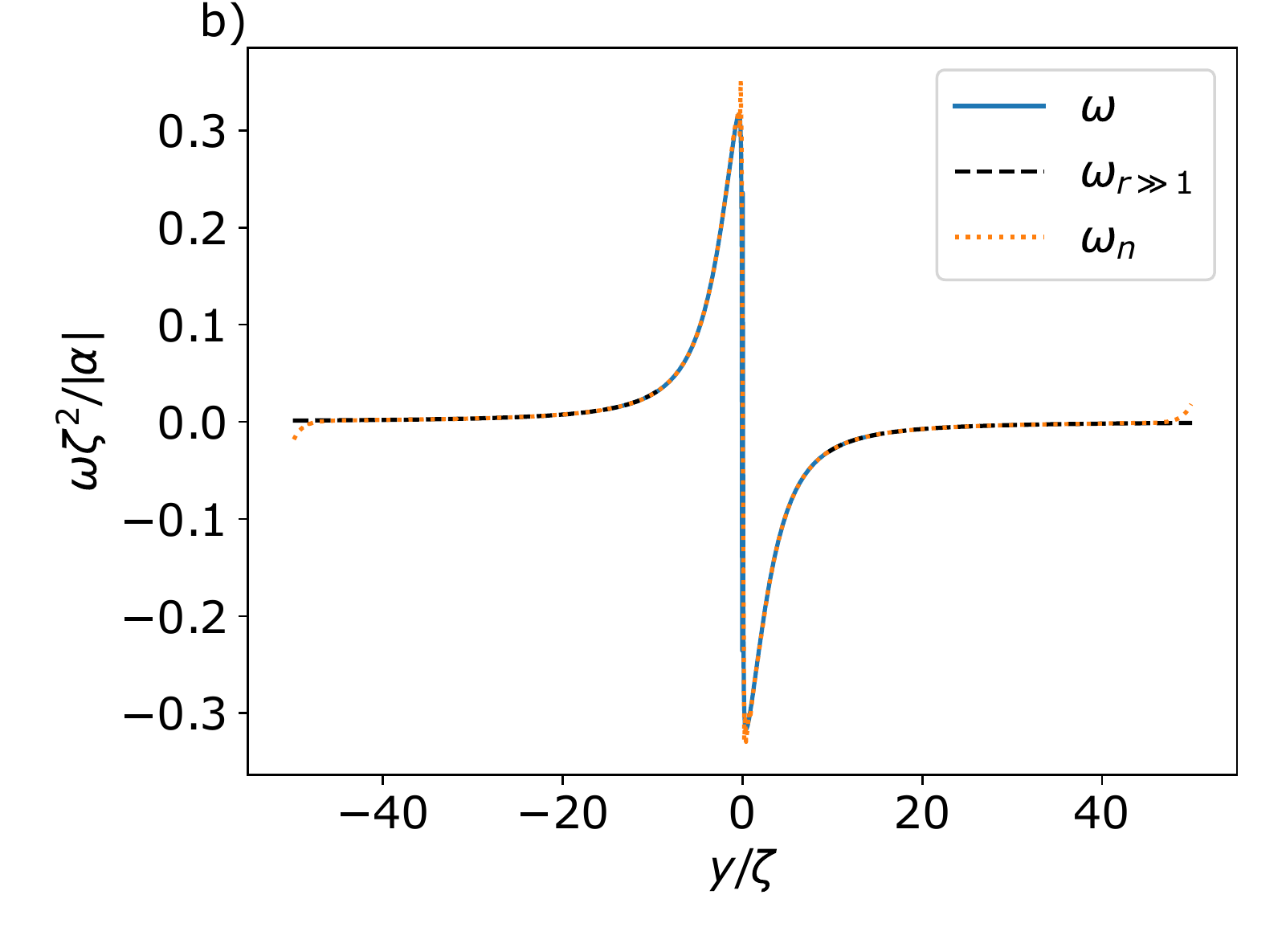}
    \includegraphics[width = 0.48\textwidth]{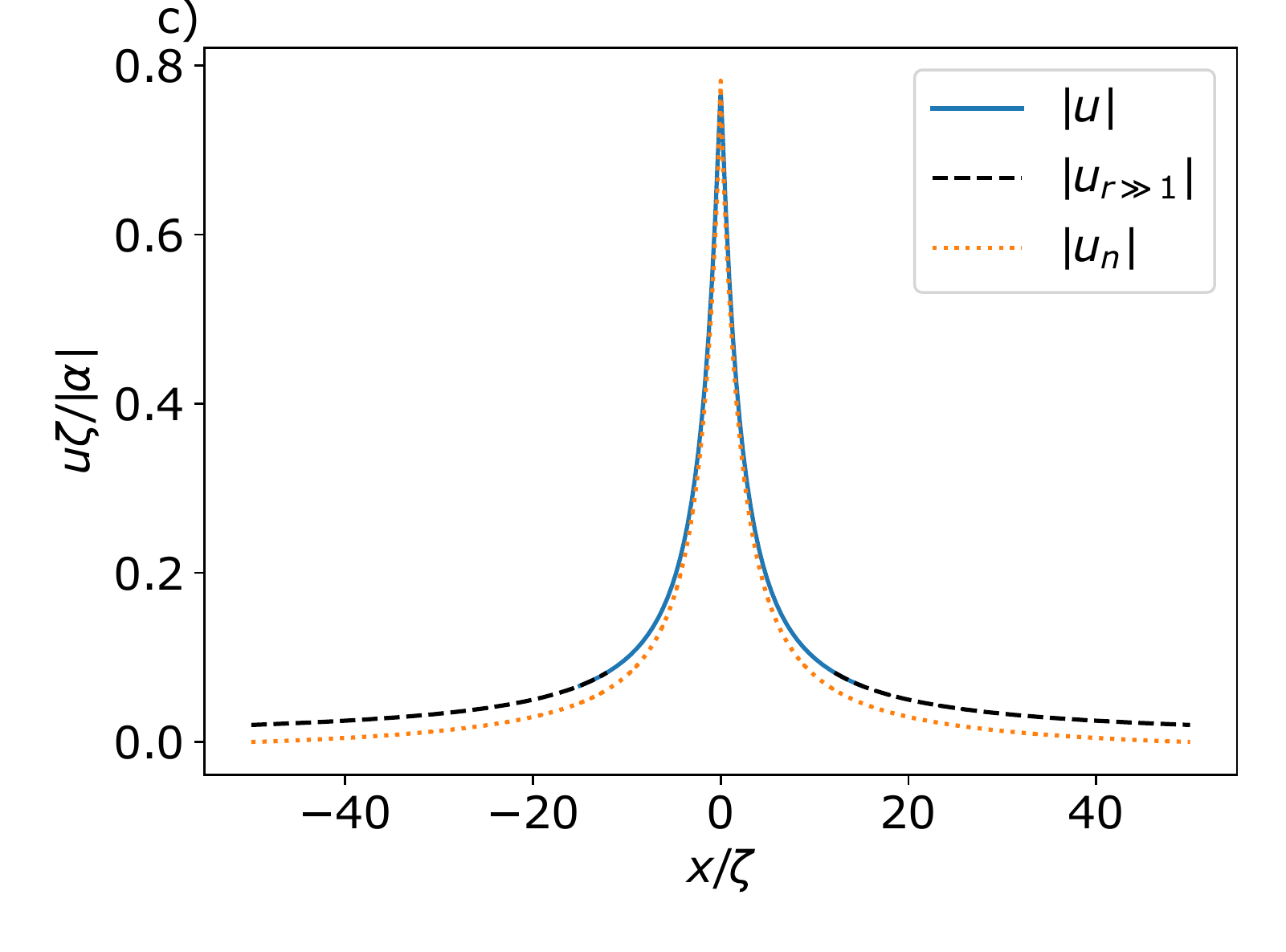}
    \includegraphics[width = 0.48\textwidth]{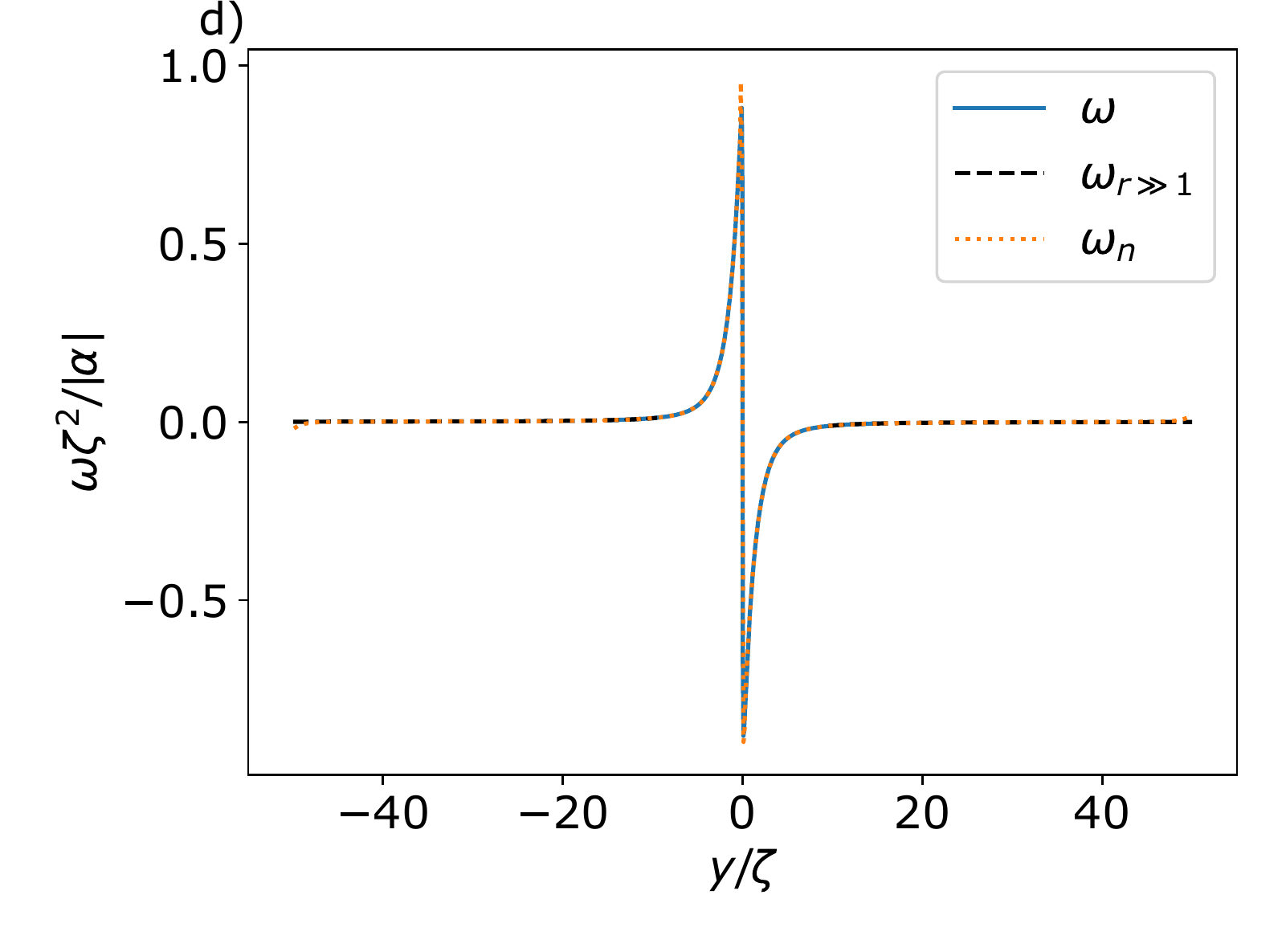}
    \caption{Cross sections of the numerically obtained velocity (a,c) and vorticity (b,d) profiles for $R= 50$ (in units of $\ell_d$) (orange dotted lines) are compared to the analytical solution for an infinite system (solid blue lines) for a negative (a,b) and positive (c,d) defect. $x$ and $y$ are also in units of $\ell_d$. The  dashed black lines are the asymptotic solutions.}
    \label{fig:num_vs_an}
\end{figure}

In Fig.~\ref{fig:num_vs_an}, we compare the cross-sectional profiles of velocity and vorticity obtained from the numerical solution for a large system, $R=50$ ($\ell_d$) to the analytical solutions. The analytical solutions is obtained by truncating the summation in the full solution at $n=5000$ and $k=500$ up to $r=15$ and then using the asymptotic solution for $r>15$. The plots of velocity in panels (a, c) show that the numerical and analytical solutions agree very well close to the defect cores, but deviate from each other near the boundary. This is due to the imposed boundary conditions of vanishing velocity field. The vorticity in Panels (b, d) agrees well in the entire domain, with a small boundary effect due to vanishing velocity and vortices spanning the system size. This effect is perhaps more visible for the negative defect and decreases with increasing $R$.

%%%%%%%%%%%%%%%%%%%%%%%%%% fig 5 %%%%%%%%%%%%%%%
%%%%%%%%%%%%%%%%%%%%%%%%%%%%%%%%%%%%%%%%%%%%%%%%%

\begin{figure}[ht]
    \centering
    \includegraphics[width = .48\textwidth]{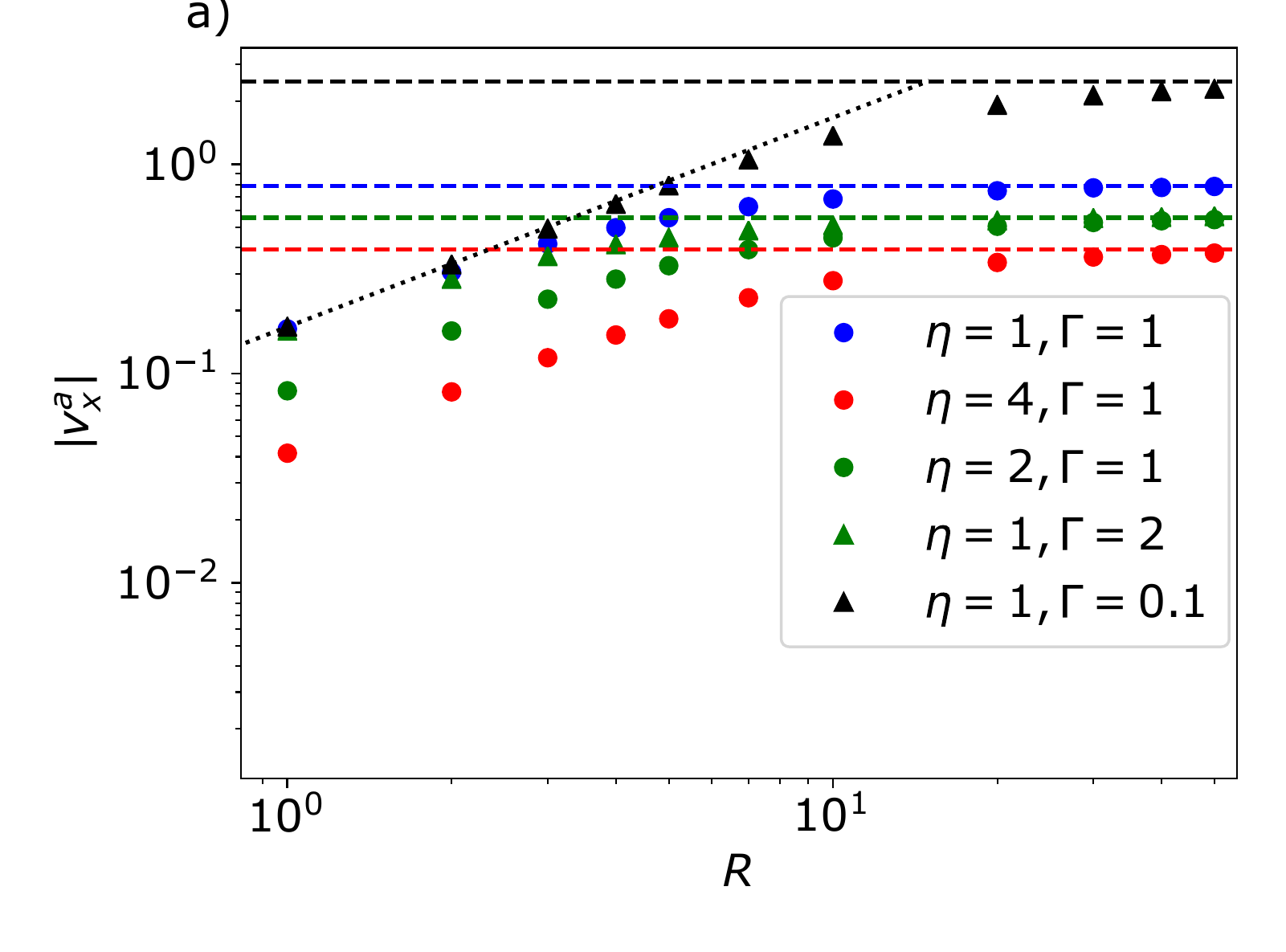}
    \includegraphics[width = .48\textwidth]{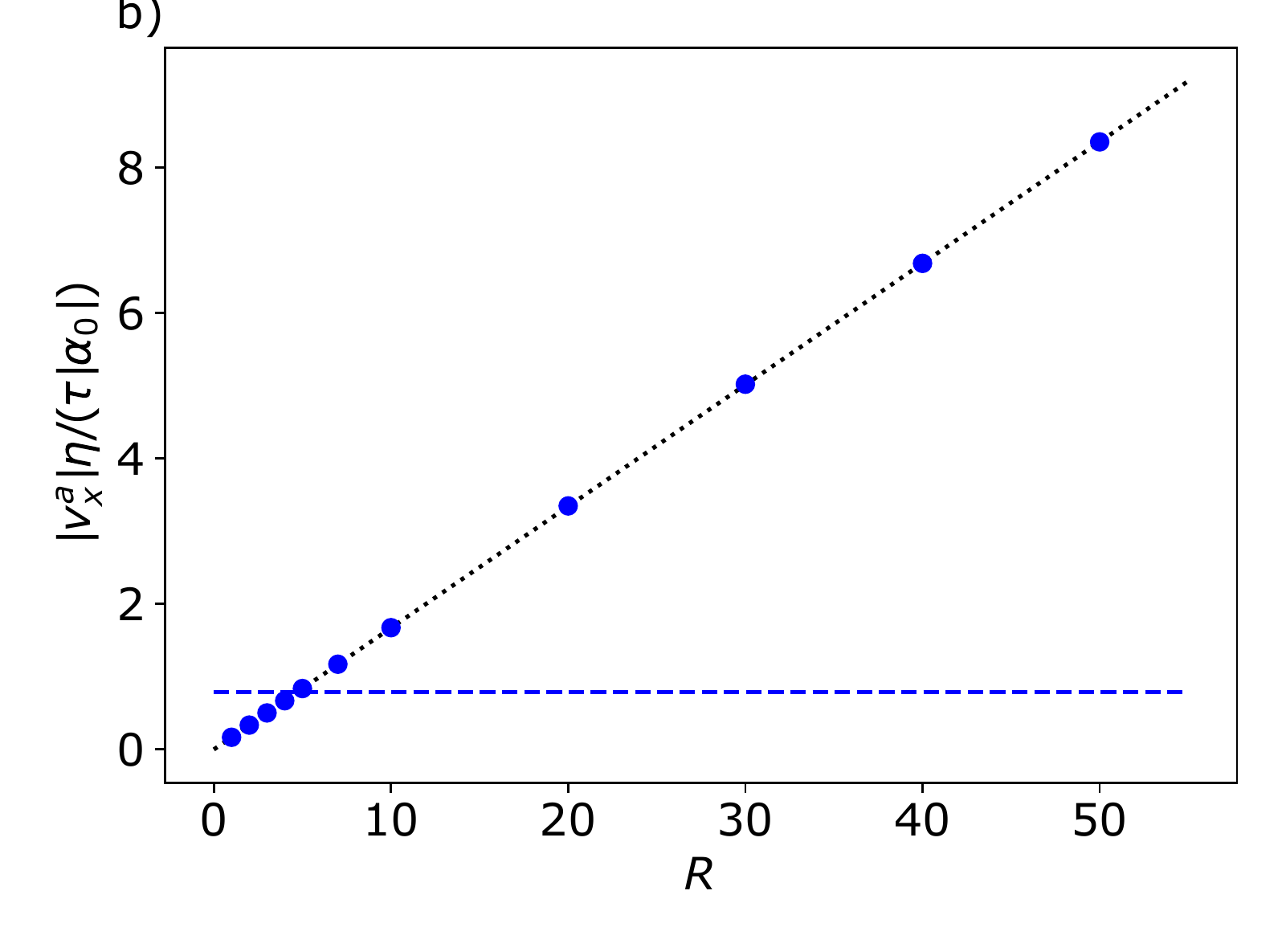}
    \caption{The  self-propulsion speed of a $+1/2$ defect as a function of the disk radius, $R$ for (a) different values of the parameters $\eta$ and $\Gamma$ and (b) $\Gamma= 0$. The black dotted line is a best fit line for the system in (b),  it is also plotted for the $\Gamma \ne 0$ systems ignoring the constant term. Horizontal dashed lines are the analytical prediction for an infinite system $v_+^{(a)} = \pi \alpha_0/(4\sqrt{\eta\tau})$ with the rescaled activity $|\alpha_0| = 1$.  
    }
    \label{fig:Numerics_vel_at_center}
\end{figure}

The self-propulsion speed $v_x^{a}$ of the $+1/2$ defect is also affected by the system size. If $\Gamma=0$ at the outset $v_x^{a}\sim R$, as noted in Ref.~\cite{giomi2014defect}. Frictional damping screens out this divergence, yielding the finite value given in Eq.~(\ref{eq:vel_at_center}) for $R\rightarrow\infty$. The numerical calculation shows, however, that for smaller $R$ there are finite-size corrections to the defect propulsive speed. These are displayed in Fig.~\ref{fig:Numerics_vel_at_center}, where we plot $|v_x^a|$ as a function of $R$ obtained from the numerical solution of Eq.~\ref{eq:Stockes} for different values of $\eta$ and $\Gamma$. The horizontal dashed lines are the analytical solution in the limit of infinite system, as given by $v_x^{(a)} \approx \frac{\pi}{4}\frac{\alpha_0}{\sqrt{\eta\Gamma}}$, while the dotted black line show the linear scaling with $R$ in the limit of zero friction. We notice that viscosity $\eta$ determines the slope for $R$ dependence in small systems,
while friction $\Gamma$ controls the cross-over to the intrinsic constant speed. Note that the asymptotic constant values of $v_x^a$ agree very well with the analytical prediction at $\zeta\gg 1$ because in the numerical computations the vortex core is actually set to zero (hydrodynamic regime with $S=1$). For comparison, we also show in Fig.~\ref{fig:Numerics_vel_at_center} (b) the defect propulsion speed in the absence of friction $\Gamma=0$ from the outset, where the speed increases linearly with the system size. The dotted black line represents the analytical prediction as found in Ref.\cite{giomi2014defect}. 

%%%%%%%%%%%%%%%% fig 6 %%%%%%%%%%%%%%
%%%%%%%%%%%%%%%%%%%%%%%%%%%%%%%%%%%

%%%%%%%%%%%%%%%%%%%%%%%%%%%%%%%%%%%%%%%%%%%%%%%%%%%%
\section{Conclusion} \label{Sec:Conclusion}
In summary, we have evaluated the flow field induced by an isolated $\pm 1/2$ defect in an incompressible active nematic film on a substrate both for an infinite system and finite-size disk.
While the self-propulsion speed of a $+1/2$ defect diverges with system size for an isolated film, we show  analytically that the  presence of finite substrate friction $\Gamma$ cures this divergence resulting in a finite speed $v_x^{(a)} \approx \frac{\pi}{4}\frac{\alpha_0}{\sqrt{\eta\Gamma}}=\frac{\pi}{4}\frac{\alpha_0}{\eta}\ell_d$ that increases with the hydrodynamic dissipation length $\ell_d$. This is also confirmed numerically in a finite disk with $R>\ell_d$. For small disks with $R<\ell_d$, the active speed scales instead linearly with $R$.  

Stable shear vortical flows are formed around the  defects. In finite systems, the size of the flow vortices is controlled by the dissipation length $\ell_d$, hence spans the whole system if $\ell_d>R$. The eye of the vortices shifts away from the defect core with increasing $R$. For infinite-size systems, the flow streamlines close at infinity as predicted by the far-field analytical solution. In the same limit, we showed that the absolute value of the velocity decreases as $1/r$ for distances that are large compared to the dissipation lengthscale, in agreement with previous studies. The $1/r$ far field decay of the flow created by defects may seem surprising as it suggests that a defect acts like a point force. This behavior arises from the long-range nature of the distortion of the texture created by defects. When other defects are present (as required in the plane to guarantee zero net topological charge), this decay is cut off by the defect separation. In finite domains it is cutoff by the system size. The $1/r$ decay indicates, however, that a multi-defect approach is needed to describe the defect gas, as attempted in Refs.~\cite{vafa2020multi,zhang2020dynamics}.

In this work we have neglected the effect of the elastic stress. An interesting extension would be to study the effects it would have on the flow field, and also considering the effect of having multiple interacting defects. \vskip6pt

%%%%%%%%%%%%%%%%%%%%%%%%%%%%%%%%%%%%%%%%%%%%%%%%%%%%%%%%%%%

\enlargethispage{20pt}

\ethics{No ethical dilemmas where encountered in the preparation of this paper. }

\dataccess{This is primarily theoretical work and does not have any experimental data. The computational data and codes for  FEniCS are available on GitHub: https://github.com/jonasron/Defect-Flows}

\aucontribute{J.R. derived the analytical results and performed finite-element
simulations for finite domains. L.A. verified all analytical calculations. All authors contributed
to a critical discussion of the analytical and numerical results and participated in writing the manuscript. L.A. conceived and coordinated the project.}

\competing{We declare we have no competing interests.}

\funding{J.R. and L.A. acknowledge support from the Research
Council of Norway through the Center of Excellence
funding scheme, Project No. 262644 (PoreLab).}

%\ack{Insert acknowledgment text here.}
%\disclaimer{No disclaimer.}

%%%%%%%%%%%%%%%%%%%%%%%%%%%%%%%%%%%%%%%%%%%%%%%%%%%%%%%
\section*{Appendix:}
\begin{appendices} 
%%%%%%%%%%%%%%%%%%%%%%%%%%%%%%%%%%%%%%%%%%%%%%%%%%%%%%

\section{ Integrals for the $+1/2$ defect}
\label{ap:Agony}
\begin{figure}
    \centering
    \includegraphics[width = .3\textwidth,trim={4cm 8cm 4cm 8cm},clip]{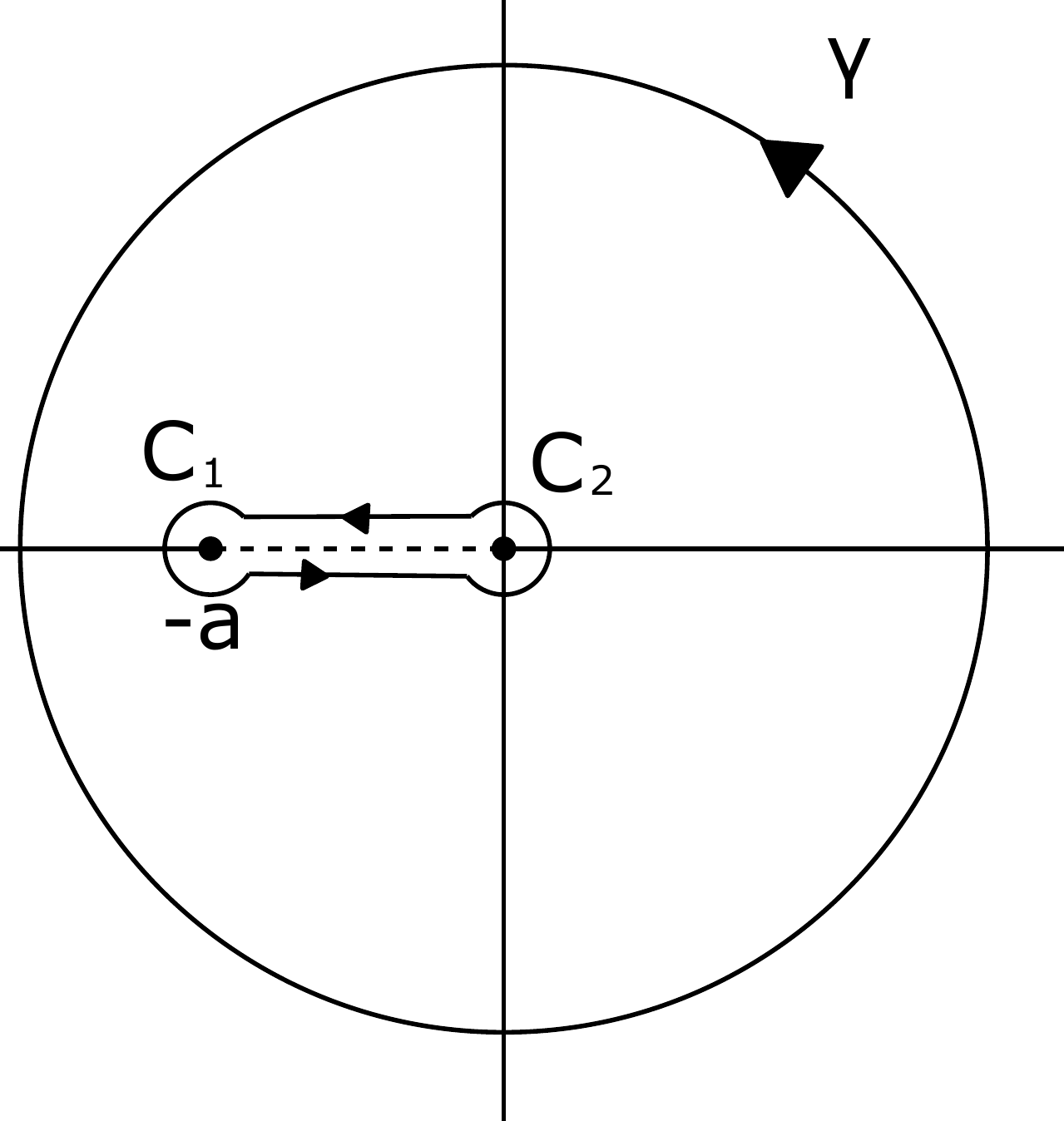}
    \caption{Sketch of the keyhole contour in the complex plane associated with the integral in Eq. \ref{eq:complex}.}
    \label{fig:1}
\end{figure}

Here, we provide the detailed steps that are taken to arrive at Eq. (\ref{eq:VelPositiveIntegralExsp}) from Eq. (\ref{eq:positive_v_integral_form}).
We start by changing to a complex representation $u = u_x + i u_y$ with complex coordinates $z = x + iy$ and $z' = x' +i y'$. By changing variables to $t = z' -z$, and then to polar coordinates $t = r' e^{i\theta} = r'\hat z$, we write Eq. (\ref{eq:positive_v_integral_form}) as

\begin{equation}
      u = \frac{ \alpha}{4 i\pi\zeta^{2} } \int d r' r'K_0( r'/\zeta ) \oint_\gamma d\hat z
    \left( \frac{r' \hat z^2+z\hat z}{( r' +\bar z\hat z) \sqrt{\hat z(r'\hat z+z)(r' +\bar z\hat z)}} + \frac{1}{\sqrt{\hat z(r'\hat z+z)(r' +\bar z\hat z)}} \right).
    \label{eq:Apandix_positive_velocity}
\end{equation}
Where $\gamma$ is the unit circle.
We notice that the integral over $\hat z$ is over three branch points.
$\hat z=0$ is always in the unit circle, 
$\hat z = -z/r$ is inside when $|z| < r$ and
$\hat z =- r/\bar z$ when $|z| > r$. 
We consider the integral over $\hat z$ and start by looking at the last term.
Splitting up the square root, we write it as
\begin{equation}
      \frac{1}{\sqrt{\bar z r'}}\oint_\gamma d\hat z\frac{1}{\sqrt{\hat z}\sqrt{(\hat z+z/r')}\sqrt{(\hat z + r'/\bar z)}}.
\end{equation}
We see that for all values of $r'$ we have two branch points inside of the contour. Therefore, we write the integral as
\begin{equation}
     \frac{1}{\sqrt{\bar z r'}}\oint_\gamma d\hat z\frac{1}{\sqrt{\hat z}\sqrt{(\hat z+a)}\sqrt{(\hat z + b)}}.
\end{equation}
From here $-a$ is the branch point inside of the contour,
while $-b$ is the point outside of the contour.
We note that the complex numbers $a$ and $b$ have the same argument $\phi$ so we can write it as
\begin{equation}\label{eq:complex}
     \frac{1}{\sqrt{\bar z r'}}\oint_\gamma d\hat z\frac{1}{\sqrt{\hat z}\sqrt{(\hat z+a e^{i\phi})}\sqrt{(\hat z + be^{i\phi})}}.
\end{equation}
$a$ and $b$ are now either $r/r'$ and $r'/r$.
If we change our integral variable from $\hat z$ to $\hat u = \hat z e^{-i\phi}$, we get
\begin{equation}
       \frac{1}{\sqrt{\bar z r'}}\oint_\gamma d\hat u e^{i\phi}\frac{1}{\sqrt{e^{i\phi}\hat u}\sqrt{(e^{i\phi}\hat u+a e^{i\phi})}\sqrt{(e^{i\phi}\hat u + be^{i\phi})}}
       = \frac{e^{-i\phi/2}}{\sqrt{\bar z r'}}\oint_\gamma d\hat u \frac{1}{\sqrt{\hat u}\sqrt{(\hat u+a )}\sqrt{(\hat u +b )}}
\end{equation}
Note that all branch points $\hat z=0$,$-a$ and $-b$ are now located on the real axis. 
We now have to consider what branch cuts we want to use to preform this integral. We consider the integral over the domain as shown in Fig. \ref{fig:1}. 
%%%%%%%%%%%%%%%%%%%%%%%% fig 7 %%%%%%%%%%%%%%%%%
%%%%%%%%%%%%%%%%%%%%%%%%%%%%%%%%%%%%%%%%%%%%%%%%
Here we have cut out a area around the branch cut in order to avoid problems. 
The key hole consists of a circle $C_1$ with radius $\epsilon$ around $-a$, the circle $C_2$ around the origin and the lines connecting them which is $\epsilon$ above or below the real line as shown in Fig. \ref{fig:1}.
Since there are no poles in the domain between the two contours, the integral of them has to be the same \cite{Kreyzig}.
The contour integral becomes
\begin{align}
    &\int_\gamma \frac{e^{-i\phi/2}}{\sqrt{\bar z r'}} d\hat u \frac{1}{\sqrt{\hat u}\sqrt{(\hat u+a )}\sqrt{(\hat u +b )}} \nonumber
    \\
    &=  \left( \int_{-a -i\epsilon\rightarrow -i\epsilon} + \int_{i\epsilon\rightarrow -a+i\epsilon} +  \int _{C_1} +\int_{C_2} \right) \frac{e^{-i\phi/2}}{\sqrt{\bar z r'}} \frac{ d\hat u}{\sqrt{\hat u}\sqrt{(\hat u+a )}\sqrt{(\hat u +b )}}. 
\end{align}
When $\epsilon \rightarrow 0$, the integrals over $C_1$ and $C_2$ disappears.
The integral above the real line is just above the branch cut and therefore positive, while the one below is negative. We therefore get
\begin{equation}
       \oint_\gamma \frac{e^{-i\phi/2}}{\sqrt{\bar z r'}} d\hat u \frac{1}{\sqrt{\hat u}\sqrt{(\hat u+a )}\sqrt{(\hat u +b )}} =- 2\frac{e^{-i\phi/2}}{\sqrt{\bar z r'}}
       \int_{-a}^0 d\hat u \frac{1}{\sqrt{\hat u}\sqrt{(\hat u+a )}\sqrt{(\hat u +b )}}.
       \label{eq:A7}
\end{equation}
Evaluating this integral, we obtain
\begin{equation}
    \oint_\gamma \frac{e^{-i\phi/2}}{\sqrt{\bar z r'}} d\hat u \frac{1}{\sqrt{\hat u}\sqrt{(\hat u+a )}\sqrt{(\hat u +b )}} =  4 i\frac{e^{-i\phi/2}}{\sqrt{\bar z r'}} \frac{1}{\sqrt{b}} K\left(\frac{a}{b}\right).
\end{equation}
Here $K$ is the complete elliptic integral of the first kind with the power series 
\begin{equation}
    K(x) = \frac{\pi}{2} \sum_{n=0}^\infty \left(\frac{(2n -1)!!}{(2n)!!}\right)^2 x^n.
\end{equation}
With the double factorial $(2n)!!= 2\cdot 4 \cdot 6 ... (2n-2) \cdot 2n$. 

We now consider the other integral over $\hat z$ in Eq. (\ref{eq:Apandix_positive_velocity}):
\begin{equation}
    \oint_\gamma d\hat z
    \frac{r' \hat z^2+z\hat z}{( r' +\bar z\hat z) \sqrt{\hat z(r'\hat z+z)(r' +\bar z\hat z)}}
    \label{eq:1/2 and -3/2 root}
\end{equation}
This is not symmetric in the branch points as the other integral was.
We can write this as
\begin{equation}
 r'^{1/2} \bar z^{-3/2}\oint_\gamma d\hat z\frac{ \hat z^{1/2} (\hat z+e^{i\phi}r/r')^{1/2}}{  (\hat z +e^{i\phi}r' /r)^{3/2}},
\end{equation}
and we can rotate the integral variable to $u =\hat z e^{-i\phi}$ and get
\begin{equation}
     r'^{1/2} \bar z^{-3/2}\oint_\gamma d\hat u e^{i\phi}\frac{ \hat u^{1/2} e^{i\phi/2} (\hat u e^{i\phi}+e^{i\phi}r/r')^{1/2}}{  (\hat u e^{i\phi} +e^{i\phi}r' /r)^{3/2}} = r'^{1/2} \bar z^{-3/2} e^{i\phi/2} \oint_\gamma d\hat u \frac{ \hat u^{1/2} (\hat u +r/r')^{1/2}}{ (\hat u  +r' /r)^{3/2}}.
\end{equation}
Which of the branch points that are inside of the integral depends on the value of $r'$. 
If $r' > r$ then $r/r'$ is inside of the units circle. 
We can then use the same branch cuts and arguments as before and find
\begin{equation}
    r'^{1/2} \bar z^{-3/2} e^{i\phi/2} \oint_\gamma d\hat u \frac{ \hat u^{1/2} (\hat u +r/r')^{1/2}}{ (\hat u  +r' /r)^{3/2}} = -2  r'^{1/2} \bar z^{-3/2} e^{i\phi/2} \int_{-r/r'}^0 d\hat u \frac{ \hat u^{1/2} (\hat u +r/r')^{1/2}}{ (\hat u  +r' /r)^{3/2}}.
\end{equation}
We can preform the integral and find that it equals
\begin{equation}
    -2  r'^{1/2} \bar z^{-3/2} e^{i\phi/2} \frac{2i }{\sqrt{r'/r}} [(2 r'/r - r/r')K(r^2/r'^2) - 2r'/r E(r^2/r'^2) ]
\end{equation}
Here $E(x)$ is the complete elliptic integral of the second kind with the power series 
\begin{equation}
    E(x) = \frac{\pi}{2} \left(1 - \sum_{n=1}^{\infty} \left(\frac{(2n-1)!!}{(2n)!!}\right)^2 \frac{x^n}{2n-1} \right).
\end{equation}
The other possibility is that $r' < r$. 
Now it is the $-3/2$ root that is inside of the integral domain.
In this case we cannot use the above approach since the integrand does not go to zero when $\hat z$ goes to $r'/r$.
However we can use the binomial expansion that is valid for $|x| < 1$.
We have that
\begin{equation}
    r'^{1/2} \bar z^{-3/2} e^{i\phi/2} \oint_\gamma d\hat u \frac{ \hat u^{1/2} (\hat u +a)^{1/2}}{ (\hat u  +b)^{3/2}}.
\end{equation}
Doing a change of variables to $h = \sqrt{u}$ with $dh = du/(2 \sqrt{u})$.
We then get (remembering a factor half since because integrating once around $h$ we have gone twice around $u$)
\begin{equation}
    r'^{1/2} \bar z^{-3/2} e^{i\phi/2} \oint_\gamma d h \frac{ h^2 (h^2 +a)^{1/2}}{ (h^2  +b)^{3/2}}.
\end{equation}
We have that $|h| = 1$, $a> 1$ and $b < 1$. 
We therefore write this
\begin{equation}
    r'^{1/2} \bar z^{-3/2} e^{i\phi/2} \oint_\gamma d h \frac{ \sqrt{a} (1 +h^2/a )^{1/2}}{ h (1  +b/h^2)^{3/2}}.
\end{equation}
We now use the binomial expansion
\begin{equation}
    (1 +x)^r = \sum_{n=0}^{\infty} {r \choose n} x^{n},
\end{equation}
and the integral becomes
\begin{equation}
    r'^{1/2} \bar z^{-3/2} e^{i\phi/2} \sqrt{a}  \oint_\gamma \frac{dh}{h} \left(\sum_{n=0}^{\infty} {1/2 \choose n} \left(\frac{h^2}{a}\right)^n \right) \left(\sum_{k=0}^{\infty} {-3/2 \choose k} \left(\frac{b}{h^2}\right)^k \right).
\end{equation}
 By the residue theorem only the terms where $k = n$ is contributing to the integral.
We therefore get
\begin{align}
    r'^{1/2} \bar z^{-3/2} e^{i\phi/2} \sqrt{a}  \oint_\gamma \frac{dh}{h}
    \left(\sum_{n=0}^{\infty} {1/2 \choose n}{-3/2 \choose n} \left(\frac{b}{a}\right)^n  \right) \nonumber\\
    = 2\pi i r'^{1/2} \bar z^{-3/2} e^{i\phi/2} \sqrt{a}  
    \left(\sum_{n=0}^{\infty} {1/2 \choose n}{-3/2 \choose n} \left(\frac{b}{a}\right)^n  \right) 
\end{align}
Inserting the expressions for $a$ and $b$ we find
\begin{equation}
    2\pi i \bar z^{-3/2} e^{i\phi/2} \sqrt{r}  
    \sum_{n=0}^{\infty} {1/2 \choose n}{-3/2 \choose n} \left(\frac{r'}{r}\right)^{2n}.  
\end{equation}
Now we have solved the integral over the angles $\hat z$. 
Inserting this into the Eq. (\ref{eq:Apandix_positive_velocity}), we find
\begin{align}
    u = \frac{\alpha}{4i\pi \zeta^{2}} \int_0^{r} dr' r' K_0(r'/\zeta)
    \Bigg(
     2\pi i \bar z^{-3/2} e^{i\phi/2} \sqrt{r}  
    \sum_{n=0}^{\infty} {1/2 \choose n}{-3/2 \choose n} \left(\frac{r'}{r}\right)^{2n} \nonumber\\
    +
    \frac{4 i e^{-i\phi/2}}{\sqrt{\bar z r'}} \sqrt{\frac{r'}{r}} K\left(\frac{r'^2}{r^2} \right)
    \Bigg)
    \nonumber
    \\
   + \frac{\alpha}{4i\pi \zeta^{2}} \int_r^{\infty} dr' r' K_0( r'/\zeta) \Bigg( \frac{4ie^{i\phi/2}\sqrt{r} }{\bar z^{3/2}} \left[ \left(  \frac{r}{r'}-2\frac{r'}{r} \right) K\left(\frac{r^2}{r'^2}\right) + 2\frac{r'}{r} E\left(\frac{r^2}{r'^2} \right)\right] \nonumber\\
   + \frac{4ie^{-i\phi/2}}{\sqrt{\bar z r'}} \sqrt{\frac{r}{r'}} K\left(\frac{r^2}{r'^2}\right) \Bigg)
\end{align}
Using that $\bar z = r e^{-i\phi}$ and 
inserting the expressions for the $K$ and $E$ this becomes
\begin{align}
      u = \frac{\alpha}{2 \zeta^{2} } \int_0^{r} dr'  K_0( r'/\zeta)\left(\frac{r'}{r}\right)^{2n+1}
      \left(
       e^{2i\phi}  
    \sum_{n=0}^{\infty} {1/2 \choose n}{-3/2 \choose n}
    + 
      \sum_{n=0}^{\infty} \left( \frac{(2n -1)!!}{(2n)!!} \right)^2 
    \right) \nonumber
    \\
    +
    \frac{\alpha}{2 \zeta^{2}} \int_r^{\infty} dr'  K_0( r'/\zeta)\left(\frac{r}{r'} \right)^{2n } \Bigg( \sum_{n=0}^\infty\left[ \left( \frac{(2n -1)!!}{(2n)!!} \right)^2 -2 \left( \frac{(2n +1)!!}{(2n+2)!!} \right)^2 \frac{2n + 2}{2n +1} \right]  e^{2i \phi} \nonumber 
    \\
    +\sum_{n=0}^{\infty} \left( \frac{(2n -1)!!}{(2n)!!} \right)^2 
    \Bigg).
\end{align}
We can simplify this by using that
\begin{equation}
      \sum_{n=0}^\infty \left( \frac{(2n -1)!!}{(2n)!!} \right)^2 -2 \left( \frac{(2n +1)!!}{(2n+2)!!} \right)^2 \frac{2n + 2}{2n +1}
      =
     - \sum_{n=0}^\infty \left( \frac{(2n -1)!!}{(2n)!!} \right)^2 \frac{n}{n+1},
\end{equation}
and for the binomial
\begin{equation}
    {1/2 \choose n}{-3/2 \choose n}  = - \frac{2n+1}{2n-1} \left( \frac{(2n-1)!!}{(2n)!!}\right)^2.
    \label{eq:binomial1}
\end{equation}
Inserting this we finally arrive at
\begin{align}
      u = \frac{\alpha}{2\zeta^{2} }
      \sum_{n=0}^{\infty}\left( \frac{(2n -1)!!}{(2n)!!} \right)^2\left(
        1 -
       \frac{2n+1}{2n-1}  e^{2i\phi}  
    \right) 
    \int_0^{r} dr'  K_0( r'/\zeta) \left(\frac{r'}{r} \right)^{2n+1 }
    \nonumber
    \\
    +
    \frac{\alpha}{2 \zeta^{2}}\sum_{n=0}^\infty \left( \frac{(2n -1)!!}{(2n)!!} \right)^2 \left(1 -\frac{n}{n+1} e^{2i \phi} 
    \right)  \int_{r}^{\infty} dr'  K_0( r'/\zeta) \left(\frac{r}{r'} \right)^{2n }.
\end{align}
%%%

%%%%%%%%%%%%%%%%%%%%%%%%%%%%%%%%%%%%%%%%%%%%%%%%%%%%%
\section{ Integrals over the Bessel function}
\label{ap:Truble}
%%%%%%%%%%%%%%%%%%%%%%%%%%%%%%%%%%%
We start a change of variables to $r'' = \zeta^{-1} r'$, and then evaluate the integrals over the Bessel functions as 
\begin{multline}
    \zeta^{2n+2}\int_0^{r/\zeta} dr''  K_0(r'') r''^{2n+1 } = \frac{1}{4}  r^{2n+2} n! \Bigg[n! {}_2\tilde F_3\left(1+n,1+n; 1,2+n,2+n; \frac{r^2}{4\zeta^{2}}\right)  
    \\
    - 2 {}_1\tilde F_2\left(1+n; 1,2+n; \frac{r^2}{4\zeta^{2}}\right) \left(\gamma + \ln\left(\frac{r}{2\zeta}\right)\right)
    +2\frac{\partial}{\partial a_1}{}_2\tilde F_3\left(a_1,1+n; 1,1,2+n; \frac{r^2}{4\zeta^{2}}\right)|_{a_1=1} 
    \Bigg]
    \\
    = \frac{1}{4}   r^{2n+2} F^{r'<r}\left(n, r/\zeta  \right),
    \label{eq:Bessel_integral_1}
\end{multline}
and
\begin{align}
    &\zeta^{1-2n}\int_{ r/\zeta}^{\infty} dr'  K_0(r') \left(\frac{1}{r'} \right)^{2n}
    \nonumber\\
    &= 
    \frac{1}{4} r^{-2n +1}
    \Gamma\left(\frac{1}{2}-n \right)\Bigg[
    4^{-n} \Gamma\left(\frac{1}{2}-n \right) \left( 2 \left(\frac{r}{\zeta}\right)^{2n-1} -4^n {}_2\tilde F_3\left(\frac{1}{2}-n,\frac{1}{2}-n; 1,\frac{3}{2}-n,\frac{3}{2}-n; \frac{ r^2}{4\zeta^2}\right)\right) \nonumber\\
    &+2  {}_1\tilde F_2\left(\frac{1}{2}-n; 1,\frac{3}{2}-n; \frac{r^2}{4\zeta^2}\right) \left(\gamma + \ln\left(\frac{ r}{2\zeta}\right)\right) 
    -2\frac{\partial}{\partial a_1}{}_2 \tilde F_3\left(a_1,\frac{1}{2}-n; 1,1,\frac{3}{2}-n; \frac{ r^2}{4\zeta^2}\right)|_{a_1 =1} \Bigg]
    \nonumber\\
    &=  \frac{1}{4} r^{-2n +1}
    F^{r' > r}\left(n, r/\zeta \right).
    \label{eq:Bessel_integral_2}
\end{align}
Here the regularized hypergeometric function is defined as 
\begin{equation}
    {}_p\tilde F_q(a_1,..,a_p;b_1...,b_q;x) = \frac{1}{\Gamma(b_1)...\Gamma(b_q)} \sum_{k=0}^{\infty} \frac{(a_1)_k...(a_p)_k}{(b_1)_k...(b_q)_k} \frac{x^k}{k!}.
\end{equation}
Where $ (a)_k = a\cdot(a+1)...(a+k-1)$ is the rising factorial.
These expressions can be simplified.
Using that $\Gamma(x) = (x-1)\Gamma(x-1)$, it follows that the Pochhammer symbol is given as
\begin{equation}
    (a)_k = \frac{\Gamma(a+k)}{\Gamma(a)},
\end{equation}
with the derivative
\begin{equation}
    \partial_a (a)_k|_{a=1} = (a)_k (\psi^{(0)}(a+k) - \psi^{(0)}(a))|_{a=1} = k! (\psi^{(0)}(1+k) -\psi^{(0)}(1)).
\end{equation}
Here $\psi^{(0)}(k)$ is the digamma function, that is the first derivative of the logarithm of the gamma function. For integer arguments, it is given as 
\begin{equation}
    \psi^{(0)}(n) = -\gamma + \sum_{l=1}^{n-1} \frac{1}{l}. 
\end{equation}
Using the relations above, we find after some algebra that the moments are given by the power series
\begin{equation}
     F^{r'<r}\left(n, r/\zeta\right) 
    = 
    \sum_{k=0}^\infty \left[ \frac{1}{(n+k+1)} -2\left(\gamma + \ln\left(\frac{r}{2\zeta}\right) \right) +2 \sum_{l=1}^{k} \frac{1}{l } \right] \frac{1}{(n+k+1)(k!)^2} \left( \frac{ r}{2\zeta}\right)^{2k}
    \label{eq:BesselMoment_F_small}
\end{equation}
and
\begin{align}
     F^{r'>r}\left(n, r/\zeta \right)
   &= 4 \sum_{k=0}^\infty \left[ \sum_{l=1}^{k} \frac{1}{l} -\left(\gamma +\ln\left( \frac{ r}{2\zeta}\right) \right) - \frac{1}{2n-1-2k} \right] \frac{1}{(2n-1-2k) (k!)^2} \left(\frac{ r}{2\zeta} \right)^{2k} \nonumber\\
   &+\frac{2\pi }{((2n-1)!!)^2}  \left(\frac{r}{\zeta}\right)^{2n-1} .
   \label{eq:BesselMoment_large}
\end{align}

%%%%%%%%%%%%%%%%%%%%%%%%%%%%%%%%%%%%%

\section{Integrals for the $-1/2$ defect}
\label{ap:dreams}
%%%%%%%%%%%%%%%%%%%%%%%%%%%%%%%%%%%
Here we provide details on the calculation leading to Eq.~(\ref{eq:Integral_eq_for_hydrodynamic_neg_defect}). Using the complex representation with similar coordinate transformations as for the $+1/2$-defect, the corresponding active flow velocity, $u^-= u^-_x+i u^-_y$ induced by active stress and pressure gradient from Eq. (\ref{eq:u_1}) reads as 
\begin{equation}
      u^- = -\frac{\alpha }{4 i\pi\zeta^{2} }\int dr' r' K_0( r'/\zeta) \oint_\gamma \frac{d\hat z}{\hat z} \left( \frac{1}{r'\hat z +z} \sqrt{\frac{r'\hat z^{-1} +\bar z}{r'\hat z +z}} + \frac{r'\hat z+z }{(r'\hat z^{-1} +\bar z)^2} \sqrt{\frac{r'\hat z+z}{r'\hat z^{-1} + \bar z}} \right).
      \label{eq:negative_before_apendix}
\end{equation}

We start by looking at the integral over $\hat z$ and we will first consider the second term:
\begin{equation}
    \oint_\gamma \frac{d\hat z}{\hat z} \frac{r'\hat z+z }{(r'\hat z^{-1} +\bar z)^2} \sqrt{\frac{r'\hat z+z}{r\hat z^{-1} + \bar z}}  
    = \frac{r'^{3/2}}{\bar z^{5/2}} \oint_\gamma d\hat z \frac{\hat z^{3/2}(\hat z + z/r')^{3/2}}{(\hat z +r'/\bar z)^{5/2}}. 
\end{equation}
Using that $z = re^{i\phi}$
and changing variable to $u = e^{i\phi}\hat z$ we get
\begin{equation}
    \frac{r'^{3/2}}{\bar z^{5/2}} \oint_\gamma d\hat z \frac{\hat z^{3/2}(\hat z +e^{i\phi} r/r')^{3/2}}{(\hat z +e^{i\phi}r'/r )^{5/2}} = \frac{r'^{3/2}}{r^{5/2}} e^{4i\phi} \oint_\gamma du \frac{u^{3/2}(u + \frac{r}{r'})^{3/2}}{(u + \frac{r'}{r})^{5/2}}.
\end{equation}
This integral has three branch points on the real axis. 
Two of these points are inside the integration domain and gives troubles. 
If $r' > r$ we have the two points $u = 0$ and $u = r/r'$ inside of the unit circle. 
In this case one use the same key hole contour technique as \ref{eq:A7} and get
\begin{equation}
     \frac{r'^{3/2}}{r^{5/2}} e^{4i\phi} \oint_\gamma du \frac{u^{3/2}(u + \frac{r}{r'})^{3/2}}{(u + \frac{r'}{r})^{5/2}} = -2  \frac{r'^{3/2}}{r^{5/2}} e^{4i\phi} \int_{-r/r'}^0 du \frac{u^{3/2}(u + \frac{r}{r'})^{3/2}}{(u + \frac{r'}{r})^{5/2}}.
\end{equation}
Preforming this integral, we arrive at 
\begin{equation}
    \frac{4 i}{3} e^{4i\phi}\frac{r'}{r^2}\left( \left[ 3\left(\frac{r}{r'}\right)^2 -16 +16 \left(\frac{r'}{r}\right)^2\right] K\left(\frac{r^2}{r'^2}\right) + \left[8 - 16 \left(\frac{r'}{r}\right)^2 \right] E\left(\frac{r^2}{r'^2}\right) \right),
\end{equation}
where the functions $K$ and $E$ are defined in  appendix \ref{ap:Agony}.
Now let us look at the integral when $r'<r$.
In this case, we can not use the contour approach because the integrand diverges near the $r'/r$ pole. 
We, therefore, use the binomial expansion to evaluate this integral.
We first change the variable to $h =\sqrt{u}$.
The integral is then
\begin{equation}
    \frac{r'^{3/2}}{r^{5/2}} e^{4i\phi} \oint_\gamma du \frac{u^{3/2}(u + \frac{r}{r'})^{3/2}}{(u + \frac{r'}{r})^{5/2}}
    =   \frac{r'^{3/2}}{r^{5/2}} e^{4i\phi} \oint_\gamma dh \frac{h^4(h^2 +a )^{3/2}}{(h^2 +b)^{5/2}}.
\end{equation}
We have introduced $a = \frac{r}{r'}$ and $b = \frac{r'}{r}$
and $|h| = 1$, $|b| < 1$ and $|a| > 1$. 
We therefore write this as
\begin{equation}
    \frac{r'^{3/2}}{r^{5/2}} a^{3/2} e^{4i\phi} \oint_\gamma dh \frac{h^4(1+ h^2/a)^{3/2}}{h^5(1 +b/h^2)^{5/2}}
    = \frac{1}{r} e^{4i\phi} \oint_\gamma \frac{dh}{h} \left( \sum_{k=0}^\infty {3/2 \choose k} \left(\frac{h^2}{a}\right)^k \right)  
    \left( \sum_{n=0}^\infty {-5/2 \choose n} \left(\frac{b}{h^2}\right)^n \right).  
\end{equation}
The residual theorem makes it so that only the terms with $k=n$ is relevant.
The integral finally becomes
\begin{equation}
    \frac{2\pi i}{r} e^{4i\phi} \sum_{n=0}^{\infty} {3/2 \choose n}{-5/2 \choose n} \left(\frac{r'}{r}\right)^{2n}.
\end{equation}
We now turn to the first term in the integral over $\hat z$ in eq. (\ref{eq:negative_before_apendix}).
It is
\begin{equation}
    \oint_\gamma \frac{d\hat z}{\hat z}  \frac{1}{r'\hat z +z} \sqrt{\frac{r'\hat z^{-1} +\bar z}{r'\hat z +z}}. 
\end{equation}
We change variables to $t = 1/\hat z$ with $dt = -d\hat z/ \hat z^2$.
In addition there comes a negative sign because we must reverse the contour. 
This integral is then
\begin{equation}
     \oint_\gamma \frac{d t}{t}  \frac{1}{r' t^{-1} +z} \sqrt{\frac{r' t +\bar z}{r' t^{-1} +z}},
\end{equation}
which is same integral as in Eq.~(\ref{eq:1/2 and -3/2 root}) with $z$ and $\bar z$ interchanged. 
We can therefore use the solution we found in appendix \ref{ap:Agony} with $\phi \rightarrow -\phi$.
The velocity field is then
\begin{align}
    u = -\frac{\alpha }{4 i\pi \zeta^{2}}\int_0^{r} dr' r' K_0( r'/\zeta) \Bigg\{ \frac{2\pi i}{r} e^{4i\phi} \sum_{n=0}^{\infty} {3/2 \choose n}{-5/2 \choose n} \left(\frac{r'}{r}\right)^{2n} 
    \nonumber\\
    + \frac{2\pi i}{r} e^{-2i\phi} \sum_{n=0}^{\infty} {1/2 \choose n}{-3/2 \choose n} \left(\frac{r'}{r}\right)^{2n}\Bigg\}
    \nonumber
    \\
    -\frac{\alpha }{4 i\pi\zeta^{2} }\int_r^{\infty} dr' r' K_0( r'/\zeta)\Bigg\{  \frac{4 i}{3} e^{4i\phi} \frac{r'}{r^2}\Bigg( \left[ 3\left(\frac{r}{r'}\right)^2 -16 +16 \left(\frac{r'}{r}\right)^2\right] K\left(\frac{r^2}{r'^2}\right) 
    \nonumber\\
    + \left[8 - 16 \left(\frac{r'}{r}\right)^2 \right] E\left(\frac{r^2}{r'^2}\right) \Bigg) 
    +4i e^{-2i\phi} \frac{1}{r}\left[ \left(\frac{r}{r'} -2\frac{r'}{r}\right) K\left(\frac{r^2}{r'^2}\right) + \frac{2r'}{r} E\left(\frac{r^2}{r'^2}\right) \right]
        \Bigg\}
\end{align}
Inserting the expressions for $K$ and $E$, and using Eq.~(\ref{eq:binomial1}) and
\begin{equation}
    \sum_{n=0}^{\infty} {3/2 \choose n}{-5/2 \choose n}
    = \frac{(2n+1)(2n+3)}{(2n-1)(2n-3)} \left( \frac{(2n -1)!!}{(2n)!!} \right)^2. 
\end{equation}
We finally arrive at
\begin{align}
    u^-(r,\phi) = &-\frac{\alpha }{2\zeta^{2} } \sum_{n=0}^{\infty} \left( \frac{(2n -1)!!}{(2n)!!} \right)^2  \frac{2n+1}{2n-1} \left[ e^{4i\phi}  \frac{2n+3}{2n-3}
    -e^{-2i\phi} \right] 
    \int_0^{r}dr' K_0(r'/\zeta) \left(\frac{r'}{r}\right)^{2n+1} \nonumber
    \\
    &-\frac{\alpha }{2\zeta^{2}}\sum_{n=0}^{\infty}\left(\frac{(2n-1)!!}{(2n)!!}\right)^2 \frac{n}{n+1}  \left[ e^{4i\phi}  \frac{n-1}{n+2} - e^{-2i\phi} \right]  \int_{r}^{\infty}dr' K_0( r'/\zeta) \left(\frac{r}{r'} \right)^{2n}.
\end{align}
%%%%%%%%%%%%%%%%%%%%%%%%%%%%%%%%%%%%%%%%%%%%
\end{appendices}

%%%%%%%%%% Insert bibliography here %%%%%%%%%%%%%%
\bibliographystyle{RS}
\bibliography{refs}

\end{document}